\documentclass[journal]{IEEEtran}
\usepackage{graphicx}
\usepackage{epstopdf}
\usepackage{epsfig}
\usepackage{subfigure}
\usepackage{multirow}
\usepackage{pifont}
\usepackage{caption}
\usepackage{tikz, pgfplots} 

\usepackage[ruled]{algorithm2e}
\usepackage{enumitem}
	
	\SetAlFnt{\small}
	\SetAlCapFnt{\small}
	\SetAlCapNameFnt{\small}
	\SetAlCapHSkip{0pt}
	\IncMargin{-\parindent}

\usepackage{cite}

\ifCLASSINFOpdf
\else
\fi
\usepackage[cmex10]{amsmath}
\usepackage{algorithmic}

\usepackage{array}

\ifCLASSOPTIONcompsoc
  \usepackage[caption=false,font=normalsize,labelfont=sf,textfont=sf]{subfig}
\else
 \usepackage[caption=false,font=footnotesize]{subfig}
\fi
\usepackage{fixltx2e}

\usepackage{stfloats}
\fnbelowfloat
\usepackage{url}

\begin{document}
\newlength\figureheight
\newlength\figurewidth
%
\title{Self-Synchronization in Duty-cycled Internet of Things (IoT) Applications}
%
%
%

\author{Poonam Yadav~\IEEEmembership{Member,~IEEE}, Julie A. McCann~\IEEEmembership{Member,~IEEE}, Tiago  Pereira
\thanks{ P. Yadav is with Computer Laboratory, University of Cambridge and J. A. McCann is with the Department of Computing, Imperial College London and T. Pereira is with the Institute of Mathematics and Computer Science, University of Sao Paulo, Brazil, e-mail: (poonam.yadav07@alumni.imperial.ac.uk).}
}

%
%

\markboth{The paper is pre-print version of IEEE INTERNET OF THINGS JOURNAL, July~2017}%
{Yadav\MakeLowercase{\textit{et al.}}: Self-Synchronization in Duty-cycled Internet of Things (IoT) Applications}
%



\maketitle

\begin{abstract}
In recent years, the networks of low-power devices have gained popularity.  Typically these devices are wireless  and interact to form large networks such as  the Machine to Machine (M2M) networks, Internet of Things (IoT), Wearable Computing, and Wireless Sensor Networks. The collaboration among these devices is a key to achieving the full potential of these networks. A major problem in this field is to guarantee robust communication between elements while keeping the whole network energy efficient. In this paper, we introduce an extended  and improved emergent broadcast slot (EBS) scheme, which facilitates collaboration for robust communication and is energy efficient. In the EBS, nodes communication unit remains in sleeping mode and are awake just to communicate. The EBS scheme is fully decentralized, that is, nodes coordinate their wake-up window in partially overlapped manner within each duty-cycle to avoid message collisions. We show the theoretical convergence behavior of the scheme, which is confirmed through real test-bed experimentation. 
\end{abstract}


%
\IEEEpeerreviewmaketitle

\section{Introduction}
%
%
%
%
\IEEEPARstart{N}{}etworks of low-power wireless embedded systems play an important role in the successful adaptation of the heterogeneous components in Internet of Things and  have gained wide attention because of their applications ranging from smart cars and cities to precision agriculture.  The end-nodes (embedded devices) of these networks are resource constrained. For example, M2M or Sensor nodes make use of low-power radios such as Bluetooth~\cite{BTnode} and Zigbee~\cite{MicaZ} for wireless communication.   These low-power radios have a limited communication range, up to 100 meters, and  data rate of a few hundred kilo-bits per second. Both design and maintenance of these networks are challenging~\cite{SunSPOT, Zhang2017} because of two major constraints: battery-power-usage and lossy communication.  
Recent work has proposed a number of solutions for designing energy efficient and robust networks. One can consider a multi-hop networking to overcome the radio's low communication range, where a source node's data is delivered to the destination node through intermediate nodes. Moreover,   duty-cycling can be used to optimize battery-power-usage and lengthen the network's lifetime~\cite{Palattella2013,Athreya2013}.  In duty-cycle networks, a node goes into a regular sleep mode  to conserve the battery power.   Combining both duty-cycled nodes and Multi-hop (DCM) networking, one can obtain an energy efficient data delivery in a large size network. The challenge is that DCM networks bring its additional overheads mainly in two ways. First,  an efficient time-synchronization scheme must be put in place to coordinate communication slot for the duty-cycled nodes.  Secondly, DCM character reduces a node's overall network throughput which is sum of node's self-generated messages, forwarded messages  and control messages. This occurs because a node has to forward  messages to its neighbors along with its self-generated messages in only a short wake-up period.  Moreover, the DCM character introduces further challenges due to the presence of unreliable wireless links and limited bandwidth. Hence, efficient route management mechanisms are central to DCM networks. 
Route management and DCM techniques aim to maximize network lifetime but at a cost. Route management mechanisms make use of control messages, which constitute extra traffic. Further, the node's duty cycling saves energy but also incurs management overheads to ensure nodes synchronize sleep-awake schedules.  
A number of \textit{centralized} time synchronization algorithms have been proposed  to achieve time coordination for low-power wireless networks~\cite{FTSP04, Yadav2007,Xu2017}. They are unsuitable for DCM networks because of issues pertaining to maintenance overheads, synchronization delays, and single points of failure~\cite{Tyrrell10}. 
The potential application scenarios of DCM networks are represented by the industrial IoT applications such as moving robots in a warehouse~\cite{Palattella2013,Daniel2013,OcadoRobot, Wan2017} or autonomous unmanned arial vehicles (drones)~\cite{Bettstetter2016} coordinate among themselves to achieve a common task.  The wireless communication nodes on robots work in duty-cycled and multi-hop networking mode to save energy consumption. In this scenario, the Emergent Broadcast Slot (EBS) scheme~\cite{Yadav2011} is suitable for achieving energy-efficiency as the scheme allows communication modules to stay in a full duty cycle to achieve synchronization among other nodes. Once such synchronization is attained,  nodes go to sleep mode and wake-up only for a short time to transmit and receive information. During the wake-up time, nodes also update their clocks, which keeps the synchronization stable and communication robust. If the quality of synchronization is compromised, the nodes then return the full duty cycle to recover synchronization. This flexibility, with no overheads, saves energy and leads to a robust scheme able to cope with dynamic changes.
		
In this paper, we present an extended version of  Emergent Broadcast Slot~(EBS) scheme~\cite{Yadav2011} that provides an energy efficient and robust communication. In our scheme  the devices stay in sleep mode and wake up in a synchronous manner for communication. This synchronization is spontaneous and requires minimal overheads.   Our synchronization protocol is biologically inspired by the firefly model and combines recent developments in the biologically inspired synchronization algorithms. Our extensive experimental and theoretical analysis reveals that the EBS has following two characteristics: \\

\begin{itemize}
\item[] --  {\it Quick Convergence and Robustness}: EBS converges exponentially fast in a fixed network structure. Moreover, in a mobile and ad-hoc network where nodes leave or join or move within the network, the scheme is robust and readapt at no significant performance cost.  \\
\item[] -- {\it Energy and Transmission Efficiency}: While keeping nodes awake only $5\%$ of their duty cycle the scheme provides $95\%$ effective message transmission. 
This saves energy and avoids message collision.  \\
\end{itemize}



The paper is organized as follows: In Section~\ref{EBS-Related}, we present related work on decentralized time synchronization schemes and firefly-based synchronization approaches. In Section~\ref{TimeSync}, we present an overview of EBS. In Section~\ref{EBSdynamics}, we present the EBS \textit{synchronization state} dynamics.
EBS configuration parameters and their effects are analyzed in Section~\ref{EBSpara}. We present two implementations of EBS in Section~\ref{EBS-Imp} and discuss their suitability and convergence in the presence of random delays. We briefly discuss our test-bed experimental setup in Section~\ref{EBS-setup}, and the resulting comparative performance results are presented in Section~\ref{EBS-Results}. We then conclude the summary along with future work in Section~\ref {EBS-Conclusion}.

\section{Related Work: Synchronization Schemes}\label{EBS-Related} 
In general, low-power nodes consist of relatively inexpensive hardware components, e.g. clocks that typically drift. Therefore, network-wide synchronization is required, thereby increasing message overheads as well as temporal and spatial instability in the network~\cite{Cheng2009,Geller2016}. Using a single central device to enforcing synchronization by dictating the time is not ideal as it imposes extra messages overload~\cite{NTP2005,Yadav2007}. Moreover, time delay for the synchronization message keeps incrementing with every hop. An additional challenge is the presence of lossy wireless links means that the synchronization packets may be dropped and would have to be sent multiple times. 

Both distributed coordination and local synchronization are used to increase stability and facilitate broadcast message transfer~\cite{Albrecht2006,Zhang2014,Yan2014}. Examples of such schemes include gradient-based~\cite{Sommer09} and bio-inspired algorithms~\cite{WernerAllen05}.  Typically bio-inspired algorithms have higher overheads than gradient-based algorithms~\cite{Bojic2014,Bojic2015,Petrocchi2016}. However, Gradient-based schemes can be ridged and unable to cope with the dynamism such as node failures~\cite{Tyrrell2010a}. On the other hand, the emergent nature of firefly-based synchronization can cope with failure. Additionally,  when a failure happens in another part of the network  firefly-based synchronization  does not affect  the local cluster.
 
\subsection{Background on Firefly Based Synchronization}
 
The pulse-coupled oscillators~(PCO) model is used to capture the synchronization behavior observed in fireflies flashing in unison \cite{mirollo1990spc}.  In this model,  each firefly is described as a periodic oscillator running at its natural frequency -- the frequency defines the number of times the firefly flashes in a unit time. And, the period defines the time interval between two consecutive flashes. 

\begin{figure}[htbp]	
    \centering
    \includegraphics[width=2in]{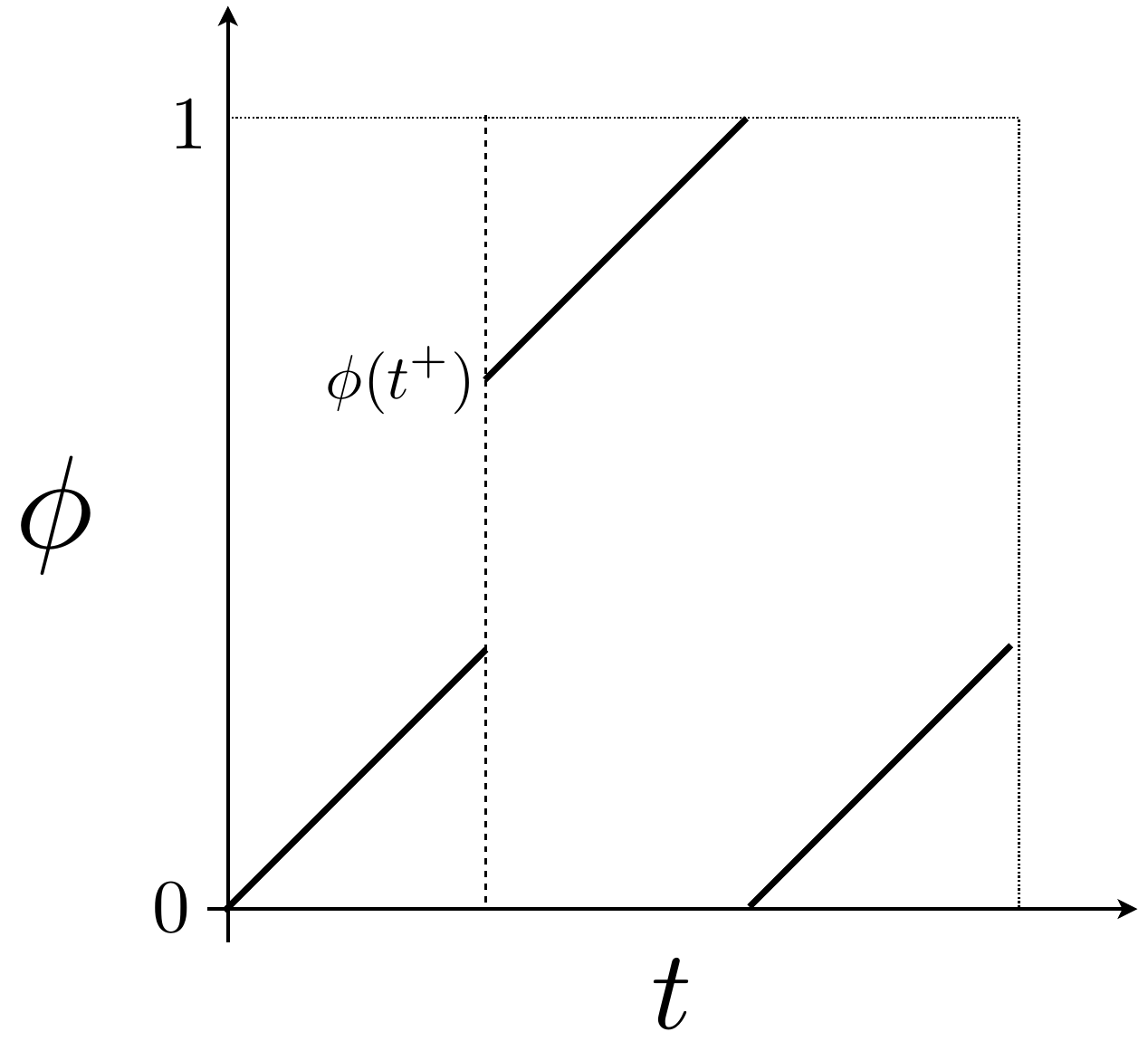} 
    \caption{ Phase Dynamics of a Pulse-Coupled Oscillator (PCO). We illustrate its behavior by showing the relationship between its phase versus time. The PCO phase increases linear with time, however, when the neighbor node fires (illustrated by the dashed line) its phase is updated to a new phase value $\phi(t^+)$. When the phase reaches $1$ it is reset to $0$.}
    \label{phase}
 \end{figure}

The state of a periodic oscillator can be described by its phase $\phi(t)$  evolving over time $t$ from $0$ to $1$.  The firefly \textit{fires or flashes} when its phase $\phi(t) = 1$ and after that resets again to $\phi(t) = 0$. To achieve synchronization, the oscillators in the system have to interact by sending and receiving pulse signals.  When a PCO's phase $\phi(t) = 1$, it emits the pulse signal (fire event) to interact with the neighbour PCOs.  Upon reception of the pulse signal, each neighbour PCO advances its phase $\phi(t)$ to a new phase $\phi(t^+)$ by taking a phase jump $\Delta \phi(t)$, which indicates the next time it will fire and is calculated by using a  \textit{phase advancement function}, see Figure~\ref{phase} for an illustration.
The different firefly based synchronization algorithms make use of different phase advancement functions designed in such a way that if all the oscillators follow a given function, they all will synchronize eventually.
 
\subsection{Related Work Based on Firefly Based Synchronization}
 Because in the PCO synchronization is an emergent property, this model has attracted much attention and has been adapted to wireless systems. For instance, variations of the model to include synchronization \textit{frustration} \cite{Closas2007} and inhibitory pulse coupling~\cite{Klinglmayr} have been introduced. An adaptive Ermentrout Model based synchronization scheme is proposed in which frequency of the flashing is adjusted  instead of phase when a node receives a flash from another node~\cite{Babaoglu2007}. 
Other relevant work includes the \textit{Reach-back} Firefly Algorithm~(RFA) that accounts for communication latencies by allowing nodes to use neighbours' firing information from the past to adjust the future firing phase \cite{WernerAllen05}.  However, due to propagation and processing delays in the notification of the firing-event, nodes keep firing which leads to chasing conditions. To avoid this, Degesys et al.~\cite{Degesys08} introduced the concept of \textit{refractory periods}. The \textit{refractory periods} are time periods that start just after node fires, and during this time, the firing node ignores fire messages from other nodes to avoid the aforementioned chasing behavior.  If the network size and connectivity are moderate  the refractory period has negligible influence on the probability of the synchronization~\cite{Wang2013}~\cite{Masood2014}. 

Another extension of the previous protocols is the extended Reachback Firefly Algorithm (e-RFA) for wireless sensor networks \cite{Leidenfrost2009}.  The synchronization algorithm uses both the \textit{Refractory Period} and the \textit{Reachback} concepts and rate calibration scheme for longer re-synchronization intervals. The algorithm is evaluated using simulations and test-bed in a mesh network topology. However, algorithm performance is presented for random multi-hop network and as well as in the presence of non-deterministic propagation delays. Implementing the RFA  in conjunction with a Late Sensitivity Window, in which nodes do not advance their phase, not only improve the percentage of  synchronized nodes, but also reduces the time to synchronize and the number of broadcast messages~\cite{cui09}.  

To reduce the number of broadcast messages and the time to synchronize an  Meshed Emergent Firefly Synchronization~(MEMFIS) has been proposed \cite{Tyrrell10}. This scheme relies on the detection of a synchronization indicator (text/symbol) common to all nodes. This indicator is embedded into each packet with payload data.  As MEMFIS embeds synchronization indicator in every payload message, it does not require separate synchronization messages but only uses payload messages as synchronization messages, which is an advantage. But embedding the synchronization indicator in every payload message also becomes a limitation when the number of network payload messages are significantly high because synchronization information consumes a significant portion of network bandwidth, lowering throughput.
 
Pagliari et al.~\cite{Pagliari2011} implemented a  scheme where an additional radio stack that is used for only synchronization messages, whereas regular CSMA radio stack for the data load. This scheme achieves fast synchronization as compared to the CSMA based schemes. However, this scheme is complex as it implements a separation of a duty-cycle into two parts for each radio stack at hardware level, which  introduces more synchronization messages collisions.

Yadav et al.~\cite{pyDcoss2011, Marcel2012} presented a receiver initiated medium access control protocol, which uses bio-inspired scheme (the initial version of EBS scheme~\cite{Yadav2011}) for route discovery messages. The paper demonstrated the extended scope of the EBS scheme in the networking stack of the wireless sensor networks and low-power wireless networks. 

In summary, bio-inspired and decentralized synchronization schemes have shown great promise. However, to be used in real-world situations,  they have to be able to work with duty-cycled systems. All current approaches ignore the fact that re-synchronization is required after a sleep period, and they also ignore the effects of asymmetric wireless links in dense networks; i.e., they are intolerant to the missing synchronization messages. Further, none of the aforementioned related schemes takes energy efficiency into account for DCM networks.

Our previous contribution~\cite{pyDcoss2011} presented an initial version of the EBS scheme showing the behavior of the scheme using the selective values of the configuration parameters.   However, it was unclear why some variable settings led to unstable behavior~ \cite{pyDcoss2011}. 
In this paper we evaluate the EBS scheme using extensive simulations and theory, thereby  deriving the relationship among different parameters  and validating the EBS performance on a real testbed.

\section{Emergent Broadcast Slot~(EBS) Scheme}\label{TimeSync}

In a typical duty-cycled multi-hop network, each node  exchanges broadcast messages with its one hop neighborhood nodes~(i.e., the nodes in its direct communication range) periodically~\cite{Yadav2009a, Yadav2009, Yadav2010}.   To minimize energy consumption, each node should remain awake for a small wake-up window within every periodic time interval $T$. Our EBS scheme provides an algorithm to achieve this small duty cycle condition. 

\begin {figure}
        \centering
  \includegraphics[width=2.5in]{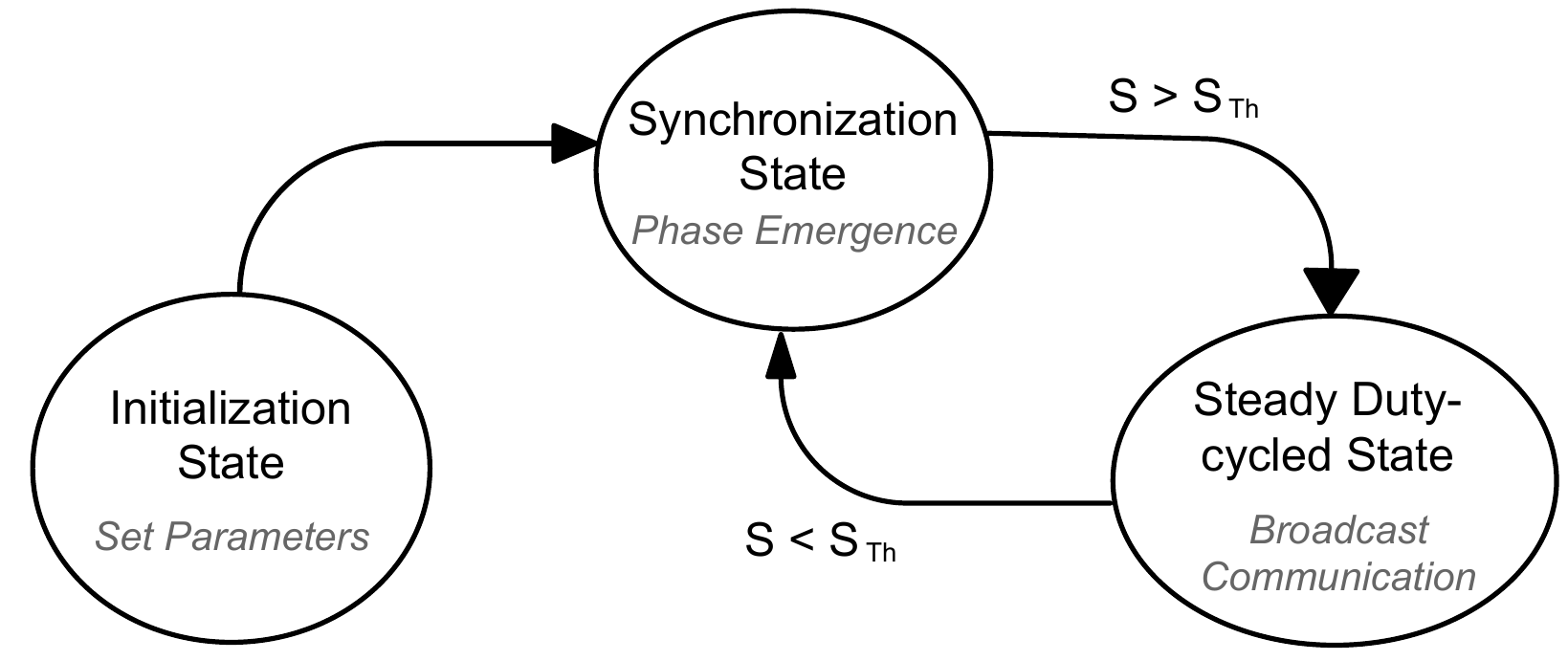}
\caption{A node is  awake  during the Initialization and Synchronization states, and in the Duty-cycled state, it wakes-up during its SETW.}
\label{fig:States}
 \end {figure}

In the EBS Scheme, each node in the network will be in one of three states: the initialization state, synchronization state, and the steady duty-cycled state.  Roughly speaking, in the initialization state the nodes are initialized and then transition to a synchronization where they will synchronize their wake-up windows; once this is achieved the nodes can go to sleep mode and only awake during the window to exchange messages. This last state is the steady duty-cycle. One important aspect of the scheme is that if the nodes feel that synchronization is compromised either by node failure or by broken connection, then they transition back to the synchronization state. In Figure~\ref{fig:States} we present a schematic representation of the stages.

\subsection{Terms and Definitions}
\begin{figure}[!t]
 \centering
 \includegraphics[width=2in]{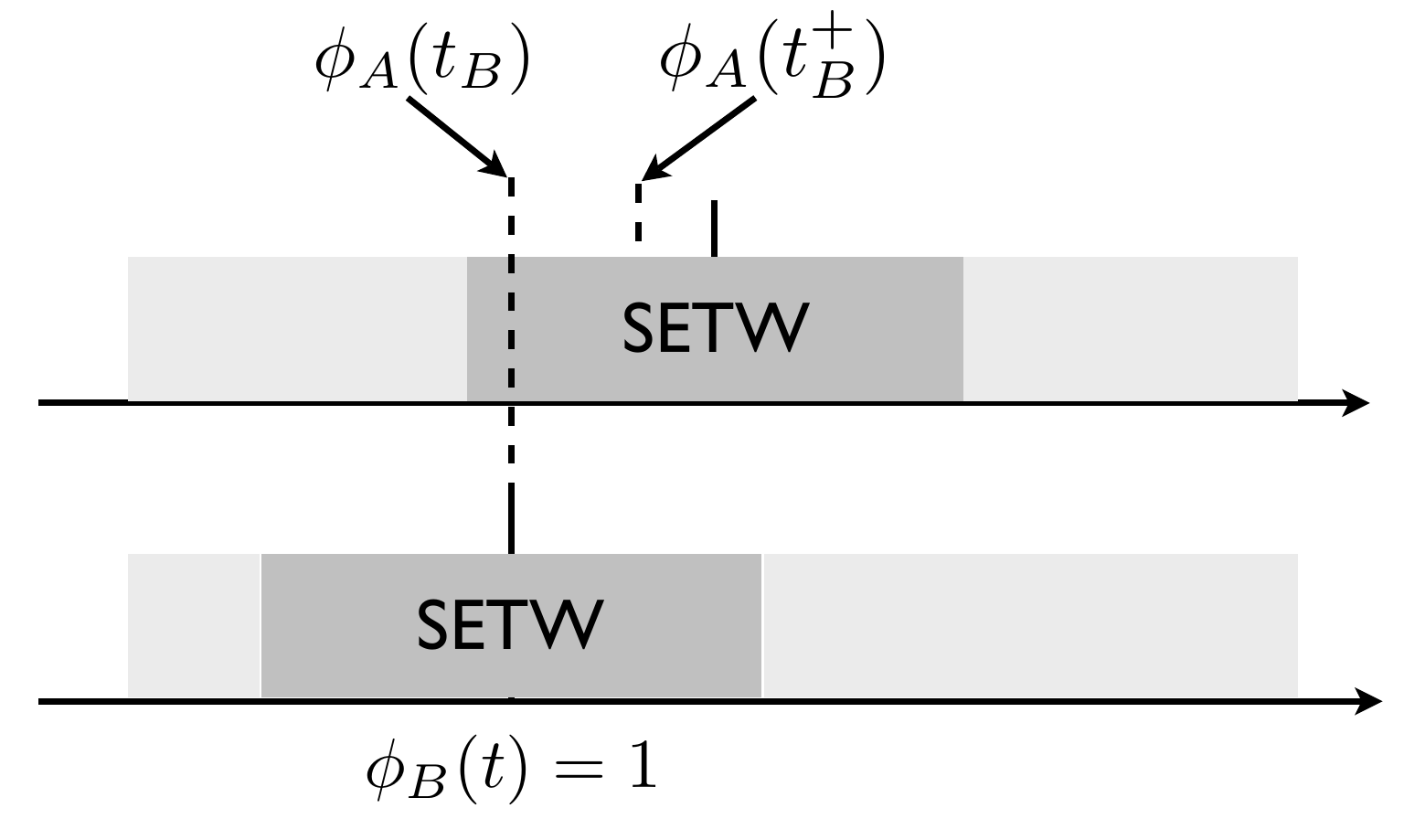} 
\caption{Phase dynamics  in the synchronization state when the node is awake for the whole period. This figure shows the phases $\phi$ and $\phi^{\prime}$ of the node A when it receives a broadcast message from B at time $t_{B}$.}
\label{fig:EBSstate}
\end{figure}

Since each node is periodic (a clock)  we can describe its state in terms of a phase variable $\phi \in \left[0,1\right]$. Recall that the node broadcast its message when $\phi$ equals $1$. We define 
$$
\phi'(t) = 1 - \phi(t)
$$ 
as the remaining phase left, after which a node is scheduled to broadcast its next control message.  See Figure~\ref{fig:EBSstate} for an illustration of the phase dynamics.

The EBS scheme makes use of the following: \\
\begin{description}
\item[Synchronization Error Tolerance Window~(SETW)] 
Given a time window $T$ and a synchronization error tolerance $\varepsilon>0$, we define
$SETW = [-\varepsilon T, \varepsilon T]$, see Figure~\ref{fig:EBSstate} for an illustration in terms of the phases. During the duty-cycle phase the nodes will only awake 
during their SETW. \\

\item[Synchronicity ($S$)]  Let $N_i$ be the set of one-hop neighbors of the node $s_i$. Moreover, let $H_i$ be the one-hop neighbors that the node $s_i$ can hear in its SETW. We define the synchronicity $S_i$ of node $s_i$ as:
$$
S_i = \frac{|H_i|}{|N_i|} \times   100 
$$  
where $| A| $ denotes the number of elements of the set $A$. \\

\item[Synchronization  Threshold ($S_{Th}$)]  A node's  $S_{Th}$ provides the minimum percentage of nodes from its neighborhood that it must hear while it is awake; i.e. during its SETW. Therefore, $S_{Th}$ will control the synchronization level we wish to enforce.
\end{description}

\subsection{Initialization State}
In the initialization state, all nodes are  awake for the  whole period, i.e., $100\%$ duty cycled. In this state the nodes start the random timer and initialize parameters such as the broadcast time
interval $T$.  Next, the nodes calculate their neighborhood size. This state lasts only a few cycles  and then transitions to the \textit{synchronization state}.

\subsection{Synchronization State}
Each node coordinate its SETW with its neighbors' using a PCO model.  The  node remains awake until their slots converge and then they sleep only to awake during the SETW.  A node declares itself synchronized\footnote{For scalability the synchronization decision is local to every node.} if its \textit{Synchronicity}, $S$ is above or equals to $S_{Th}$. Once  a node declares itself as synchronized, it transitions to the steady duty-cycled state.

To achieve the required $S$ we use a PCO model with a particular phase advancement function
\begin{equation}\label{eq2}
\phi(t^+) = 
\left\{
\begin{array}{cc}
1- g(\phi(t))  &  \mbox{ if }  \varepsilon < \phi(t) <(1-\varepsilon) \\
\phi(t)  &  \mbox{Otherwise}, 
\end{array}
\right.
\end{equation}
with  $\varepsilon \leq (0.5)$,  and $\phi(t^+)$ denotes the new phase value and $ \phi(t)$ is the previous phase value. $g$ is the phase advancement function given by:
$$
g(\phi(t)) =  \sigma(1- \phi(t)) \,\,\,\,\,\, \mbox{ with } \sigma \in (0,1).
$$ 

Suppose, node A receives a broadcast message from node B at time $t_B$ as shown in Figure~\ref{fig:EBSstate}. Then node A checks the time it has passed since its previous broadcast message in the current period, represented by $\phi_{A}(t_{B})T$. 
According to the Equation~(\ref{eq2}), if $(\varepsilon < \phi_{A}(t_{B}) < (1-\varepsilon))$, it shortens its remaining phase $\phi_{A}'(t_{B})$ to  $g(\phi_{A}(t_{B}))$.

In ideal conditions, to achieve synchronization fast,  the node transmits its broadcast message immediately  by setting $\phi_{A}(t_{B}^+) = 1$ and  $g(\phi_{A}(t_{B})) = 0$.   After transmission it resets the next broadcast message transmission time to $T$ and  $\phi_{A}(t_{B} = 0)$. However, recall that all nodes within the one-hop neighborhood can listen to the same broadcast message and if they were to broadcast their messages on receipt  of this broadcast message simultaneously, then there will be many messages in the local area; leading to packet collisions\footnote{Original CSMA back-off mechanism  introduces delays greater than the SETW which disturbs synchronization that as a result increases message losses. Therefore, in our implementation we  keep CSMA back-off and re-transmissions  to the minimum value to keep synchronization less affected, which as a result makes CSMA more effective.}. To avoid such collisions, nodes delay their broadcast message transmission by ($\phi'(t^+)$) which is equal to $g(\phi(t))$ without disturbing the synchronization. 

The appropriate values of $\varepsilon$ and $\sigma$ depend on a number of factors such as the number of nodes in a node's 1-hop and 2-hop neighborhoods, the value of $T$, as well as message propagation delays (discussed further in Section~\ref{EBSpara}).  The details of how synchronization is achieved in this state are discussed in the next Section~\ref{EBSdynamics}.

\subsection{Steady Duty-cycled State}
Once synchronized, the nodes switch to the \textit{duty-cycle phase} and wake-up only during their respective SETW to exchange messages. In the \textit{duty-cycle state}, all nodes also constantly 
update their $S$ values. The nodes whose $S$ values fall below the  $S_{Th}$, switch back to the \textit{Synchronization State} to recalculate the $S_{Th}$ for a single $T$ period by returning to $100\%$ duty-cycling and then switching to the \textit{Synchronization State}.
The various reasons for a drop in a node's $S$  value can be due to clock drift, change in network density, etc. The two common cases are:

{\bf New node joins the network}: It is the new node's responsibility to synchronize itself with the already synchronized network. First, it follows two first states gathering information regarding the number of neighbors it can hear, and then synchronizes its SETW with its neighbors. To this end, the node calculates $S$ and it switches to  the \textit{Steady Duty-cycled State} if its $S \geq S_{Th}$.  If the new joining node  is mobile and already in the \textit{Steady Duty-cycled State},  then it switches to the \textit{Synchronization State} only if $S \leq S_{Th}$.  The neighbor nodes, who are now synchronized with this node, find  their neighbors count within SETW is incremented by 1. Until synchronicity $S$ of  the neighbor nodes satisfy the condition $100\geq S \geq S_{Th}$, they continue without any change, but if $S > 100$, then nodes update their neighbors count and $S_{Th}$.

{\bf A node leaves the network}: When a synchronized node leaves the  network, it decreases the value of $S$ of its neighboring nodes as well as total number of neighbors. If this change drops their value in such way that  $S < S_{Th}$,  then nodes update their number of neighbors and $S$ by switching back to \textit{Synchronization State} otherwise they continue  without any change, however, new neighbor count will be updated only when nodes go to the  \textit{Synchronization State}.

This method of switching back has the advantage that, unlike gradient 
based schemes~\cite{Sommer09}, one node's disturbance does not perturb the whole network; therefore, EBS scheme is both agile and tolerant of change. The EBS requires the node to wake-up only once within the period $T$ rather than multiple times. Recall that multiple wake-up periods within a $T$, can cause higher-energy consumption in terms of  radio state transitions as the transceiver is required to power up, transmit and receive messages for each period. All current distributed or centralized slot-based time  synchronization schemes used in multi-hop networks expect to have multiple wake-up periods~\cite{Ye2002,SCP06MAC}.

\section{EBS Synchronization State Dynamics}\label{EBSdynamics}

We model the EBS System as a complex dynamical network represented by a graph $G(D, L)$, that is composed of $n$ nodes represented  by the set $D = (s_1, s_2,\ldots, s_n)$ where $L$ is a set of all transmission links between nodes in $D$.  The dynamics of each node $s_i$ in the network is characterized by the phase $\phi_i$. Moreover, the one-hop neighbourhood of a node $s_i$ is represented by the set $N_i$. 

We start by examining the EBS dynamics in the absence of transmission delays. EBS is required to  achieve synchronization to a given precision to avoid the message collisions that would be present\footnote{There is a  trade off between the SETW size and the probability of network collisions. A smaller SETW has energy saving advantages but leaves a smaller time interval within which messages can be communicated thus increasing collisions.}. Our goal is to study the ability of the EBS to quickly synchronize so that the number of nodes in $100\%$ duty-cycle states are minimized. To this end, we make use of two metrics: the average phase difference and the average phase advancement. The  average phase difference is given by: 
\begin{equation}\label{avgPhase}
\Delta \phi = \frac{1}{n}\sum\limits_{s_ i\in D} \frac{1}{|N_i|}\sum\limits_{s_ j \in N_i } | \phi_i -\phi_j |.
\end{equation}
Here, $\emph{$N_i$}$  represents the neighbours of node $s_i$ and the total number of neighbours is $|N_i|$. \\
The average phase advancement at time $t$ in time-period $T$ is calculated as follows:
\begin{equation}\label{avgAdvance}
\Delta^+ = \frac{1}{n}\sum\limits_{ s_ i \in D} |\phi_i(t^+ ) -\phi_i(t) |.
\end{equation}

The $\Delta^{+}$ provides the average phase displacement a node must perform. So, decreasing $\Delta^{+}$ leads to a better performance of the scheme and while achieving the required precision.

\begin{figure}[!ht]
\centering  
\includegraphics[width=1.6in]{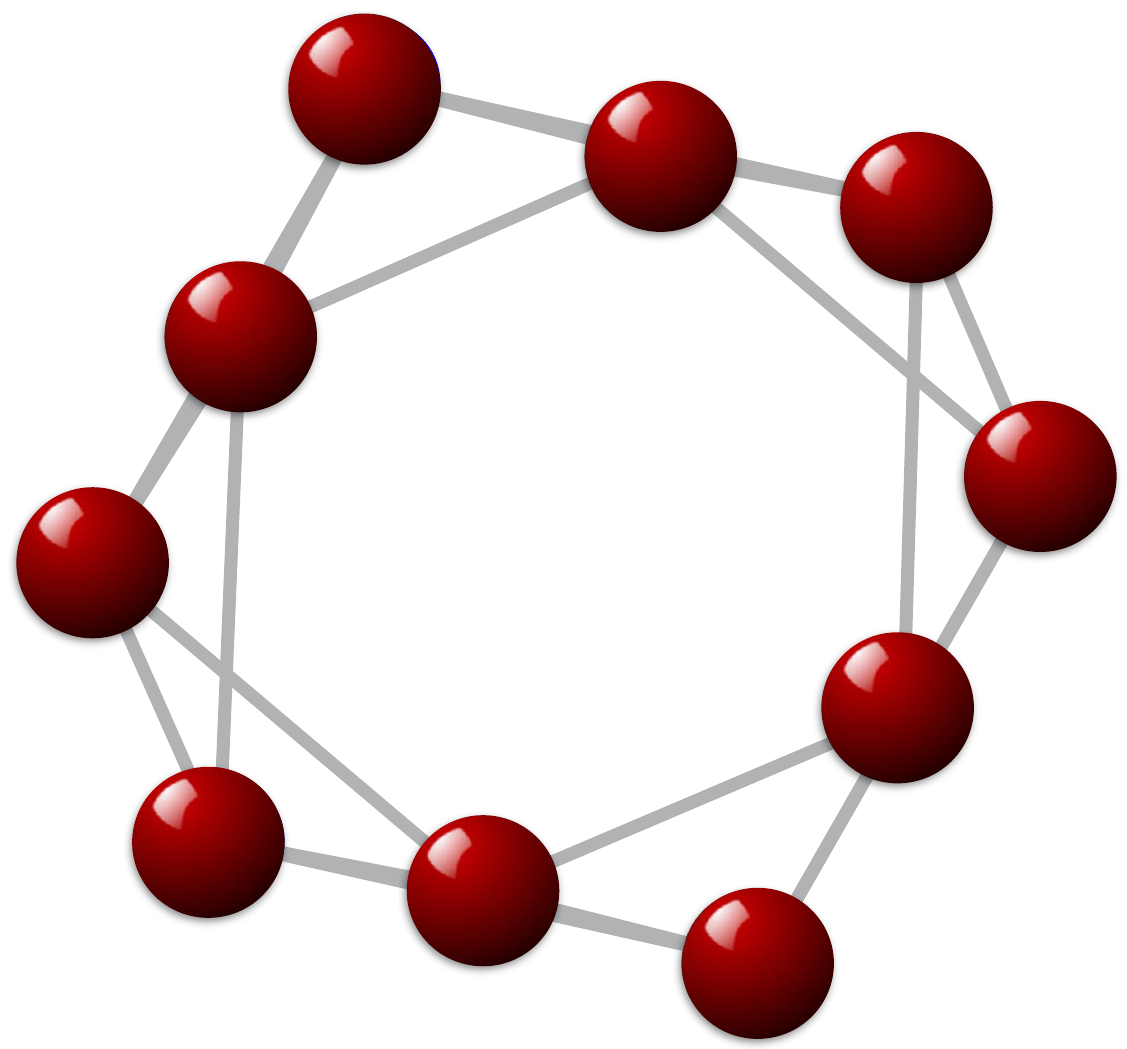}
\caption{Simulation Topology }
\label{fig:topo2}
\end{figure}

\subsection{Synchronization Error-Tolerance}\label{Precision} 
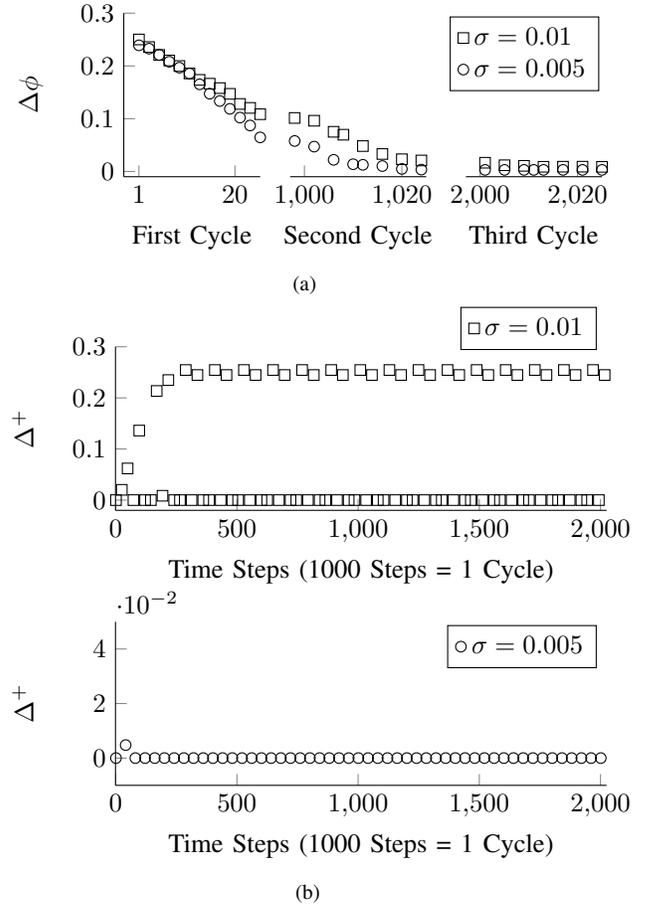
\begin{figure}[ht!] 
   \centering
   \subfigure[]{
    \setlength\figureheight{0.12\textwidth}
   \setlength\figurewidth{0.10\textwidth}
%
\begin{tikzpicture}

\begin{axis}[%
width=\figurewidth,
height=\figureheight,
scale only axis,
xmin=997,
xmax=1025,
xtick ={1000, 1020},
xlabel={Second Cycle},
every outer y axis line/.append style={white},
every y tick label/.append style={font=\color{white}},
ymin=-0.01,
ymax=0.3,
ytick={\empty},
name=plot2,
axis x line*=bottom,
axis y line*=left
]
\addplot[only marks,mark=square,mark options={},color=black] plot table[row sep=crcr,]{%
998	0.101361946532\\
1002	0.096092597332\\
1006	0.075151959112\\
1008	0.069621921576\\
1012	0.048107598202\\
1016	0.03306294683\\
1020	0.02311947492932\\
1024	0.0207765907315\\
};
\addplot[only marks,mark=o,mark options={},color=black] plot table[row sep=crcr,]{%
998	0.057731835228\\
1002	0.047044755228\\
1006	0.0220328901505\\
1010	0.013645952600752\\
1012	0.01256511461588\\
1016	0.010230226907066\\
1020	0.00477398272882\\
1024	0.003496547932648\\
};

\end{axis}

\begin{axis}[%
width=\figurewidth,
height=\figureheight,
scale only axis,
xmin=-2,
xmax=25,
xlabel={First Cycle},
xtick ={1, 20},
ymin=-0.01,
ymax=0.3,
ylabel={$\Delta \phi$},
at=(plot2.left of south west),
anchor=right of south east,
axis x line*=bottom,
axis y line*=left
]
\addplot[only marks,mark=square,mark options={},color=black] plot table[row sep=crcr,]{%
1	0.25015126\\
3	0.23579356\\
5	0.22086684\\
7	0.2102414132\\
9	0.1997984732\\
11	0.1856842932\\
13	0.1735396024\\
15	0.166319072\\
17	0.1579529516\\
19	0.1470568232\\
21	0.1278805832\\
23	0.120417459\\
25	0.108470643532\\
};
\addplot[only marks,mark=o,mark options={},color=black] plot table[row sep=crcr,]{%
1	0.23902347\\
3	0.23278981\\
5	0.2209177\\
7	0.2079327\\
9	0.19667886\\
11	0.1855759356\\
13	0.1651752056\\
15	0.1473302956\\
17	0.1334936901\\
19	0.1186093301\\
21	0.1020755518\\
23	0.0870750118\\
25	0.0645424718\\
};

\end{axis}

\begin{axis}[%
width=\figurewidth,
height=\figureheight,
scale only axis,
xmin=1997,
xmax=2025,
xlabel={Third Cycle},
every outer y axis line/.append style={white},
every y tick label/.append style={font=\color{white}},
ymin=-0.01,
ymax=0.3,
xtick ={2000, 2020},
ytick={\empty},
at=(plot2.right of south east),
anchor=left of south west,
axis x line*=bottom,
axis y line*=left,
legend style={draw=black,fill=white,legend cell align=left}
]
\addplot[only marks,mark=square,mark options={},color=black] plot table[row sep=crcr,]{%
1025	0.015999661502\\
2001	0.01649120567318\\
2005	0.011733272569082\\
2009	0.010625289152728\\
2013	0.008813459079704\\
2017	0.008970634268776\\
2021	0.009031211410664\\
2025	0.008654026619992\\
3003	0.008861660622888\\
3007	0.009086843094488\\
3011	0.009158718867888\\
};
\addlegendentry{{$\sigma = 0.01$}};

\addplot[only marks,mark=o,mark options={},color=black] plot table[row sep=crcr,]{%
1025	0.003553037811858\\
2001	0.002909838689664\\
2005	0.002907026004446\\
2009	0.002917010087462\\
2011	0.0029070100004\\
2013	0.002907010000002\\
2017	0.00290701\\
2021	0.00291701\\
2025	0.00290701\\
3003	0.00290701\\
3005	0.00290701\\
3007	0.00291701\\
3011	0.00290701\\
};
\addlegendentry{{$\sigma = 0.005$}};

\end{axis}
\end{tikzpicture}%
  }
   \subfigure[]{
     \setlength\figureheight{0.12\textwidth}
   \setlength\figurewidth{0.36\textwidth} 
%
\begin{tikzpicture}

\begin{axis}[%
width=\figurewidth,
height=\figureheight,
scale only axis,
xmin=-1,
xmax=2025,
xlabel={Time Steps (1000 Steps = 1 Cycle)},
ymin=-0.01,
ymax=0.05,
ylabel={$\Delta^{+}$},
name=plot2,
axis x line*=bottom,
axis y line*=left,
legend style={draw=black,fill=white,legend cell align=left,anchor=north east}
]
\addplot[only marks,mark=o,mark options={},color=black] plot table[row sep=crcr,]{%
1	0\\
41	0.004766\\
81	0\\
121	0\\
161	0\\
201	0\\
241	0\\
281	0\\
321	0\\
361	0\\
401	0\\
441	0\\
481	0\\
521	0\\
561	0\\
601	0\\
641	0\\
681	0\\
721	0\\
761	0\\
801	0\\
841	0\\
881	0\\
921	0\\
961	0\\
1001	0\\
1041	0\\
1081	0\\
1121	0\\
1161	0\\
1201	0\\
1241	0\\
1281	0\\
1321	0\\
1361	0\\
1401	0\\
1441	0\\
1481	0\\
1521	0\\
1561	0\\
1601	0\\
1641	0\\
1681	0\\
1721	0\\
1761	0\\
1801	0\\
1841	0\\
1881	0\\
1921	0\\
1961	0\\
2001	0\\
};
\addlegendentry{{$\sigma = 0.005$}};

\end{axis}

\begin{axis}[%
width=\figurewidth,
height=\figureheight,
scale only axis,
xmin=-1,
xmax=2025,
xlabel={Time Steps (1000 Steps = 1 Cycle)},
ymin=-0.02,
ymax=0.3,
ylabel={$\Delta^{+}$},
at=(plot2.above north west),
anchor=below south west,
axis x line*=bottom,
axis y line*=left,
legend style={draw=black,fill=white,legend cell align=left,anchor=south east}
]
\addplot[only marks,mark=square,mark options={},color=black] plot table[row sep=crcr,]{%
1	0\\
25	0.0201262\\
49	0.06189453956\\
73	0\\
97	0.136074264906\\
121	0\\
145	0\\
169	0.213523418668\\
193	0.0085238\\
217	0.234865360159\\
241	0\\
265	0\\
289	0.254539999895\\
313	0\\
337	0.24475\\
361	0\\
385	0\\
409	0.25454\\
433	0\\
457	0.24475\\
481	0\\
505	0\\
529	0.25454\\
553	0\\
577	0.24475\\
601	0\\
625	0\\
649	0.25454\\
673	0\\
697	0.24475\\
721	0\\
745	0\\
769	0.25454\\
793	0\\
817	0.24475\\
841	0\\
865	0\\
889	0.25454\\
913	0\\
937	0.24475\\
961	0\\
985	0\\
1009	0.25454\\
1033	0\\
1057	0.24475\\
1081	0\\
1105	0\\
1129	0.25454\\
1153	0\\
1177	0.24475\\
1201	0\\
1225	0\\
1249	0.25454\\
1273	0\\
1297	0.24475\\
1321	0\\
1345	0\\
1369	0.25454\\
1393	0\\
1417	0.24475\\
1441	0\\
1465	0\\
1489	0.25454\\
1513	0\\
1537	0.24475\\
1561	0\\
1585	0\\
1609	0.25454\\
1633	0\\
1657	0.24475\\
1681	0\\
1705	0\\
1729	0.25454\\
1753	0\\
1777	0.24475\\
1801	0\\
1825	0\\
1849	0.25454\\
1873	0\\
1897	0.24475\\
1921	0\\
1945	0\\
1969	0.25454\\
1993	0\\
2017	0.24475\\
};
\addlegendentry{{$\sigma = 0.01$}};

\end{axis}
\end{tikzpicture}%
  }
\caption{Convergence behavior of EBS with the different values of $\sigma$ when  $\delta = 0$  and $\varepsilon = 0.01$.  Fig (a) shows that the average phase-difference  converges to zero for  different values of $\sigma$ after few cycles.  Fig (b) shows that average phase advancement converges to zero when $\sigma = 0.005$ and stays to a constant value when  $\sigma = 0.01$, hence represents non-convergence.  It is clear from the figures that EBS achieves both required precision and stability when values of $\sigma$ and $\epsilon$ satisfy Equation~(\ref{StabC}).}
\label{fig:figure1}
\end{figure}

We wish to derive a stability condition for pairwise synchronization~\cite{Tiago2014, Tiago2013}. Consider a given node $s_i$ and let $s_j$ belong to $N_i$ that one-hop neighbourhood of $s_i$. Next, we assume that the two nodes are synchronized, that is, they fire within the $SETW$. The synchronization condition  imposes that when the node $s_i$ fires at time $t$  the new value $\phi_j(t^+)$ must be in $SETW$. This translates to :
$$
| \phi_i(t) - \phi_j(t^+)  | < \varepsilon,
$$
using Equation~(\ref{eq2}) we obtain the fact that when $s_i$ fires we have $\phi_i(t)=1$,
$$|\sigma (1-\phi_{j})| < \varepsilon, 
$$
notice that  $|1-\phi| \le 1 -\varepsilon $, hence the stability condition reads as
\begin{equation}\label{StabC}
\sigma < \frac{\varepsilon}{1 -\varepsilon}.
\end{equation}

We analyzed the convergence behavior of EBS through simulations using different network topologies. However, due to space limit we present convergence behavior for a regular network with four nearest neighbours as  shown in Figure~\ref{fig:topo2}, because it provides a good example of complex scenario  where each node has four connections with triangle cliques.  

We first analyzed the behavior of EBS for different values of $\varepsilon$ and $\sigma$ in the absence of delay. In Figure~\ref{fig:figure1}~(a), we found that for different values 
of $\sigma$, EBS reaches the required average phase difference (refer Equation~(\ref{avgPhase})) in two cycles. To better understand how well the duty cycle 
is being achieved we also analyzed the average phase advancement (refer Equation~(\ref{avgAdvance})) of the nodes over time. 
If the value of this metric is greater than $0$, then the nodes phase-jump using the EBS phase advancement 
function. In Figure~\ref{fig:figure1}~(b), we see that the average phase advancement value $\Delta^+$  in every time period $T$ reduces to $0$ after a few time steps when our stability criterion shown in Equation~(\ref{StabC})  is met. 

\subsection{Convergence in Presence of Deterministic Delays}\label{DeterDelay}
Let the phase $\delta$ represents a phase change during message propagation between any two directly connected nodes. Notice that $\delta$ represents a deterministic delay. In this case, to obtain synchronization between the SETW's we require:

\begin{equation}\label{Eq-delay}
| \phi_i(t) - \phi_j(t^+) | <  \varepsilon - 2\delta.
\end{equation}

\begin{figure}[ht!] 
 \begin{minipage}{0.9\linewidth}
   \centering
    \subfigure[]{
    \setlength\figureheight{0.3\textwidth}
   \setlength\figurewidth{0.7\textwidth}
   \input{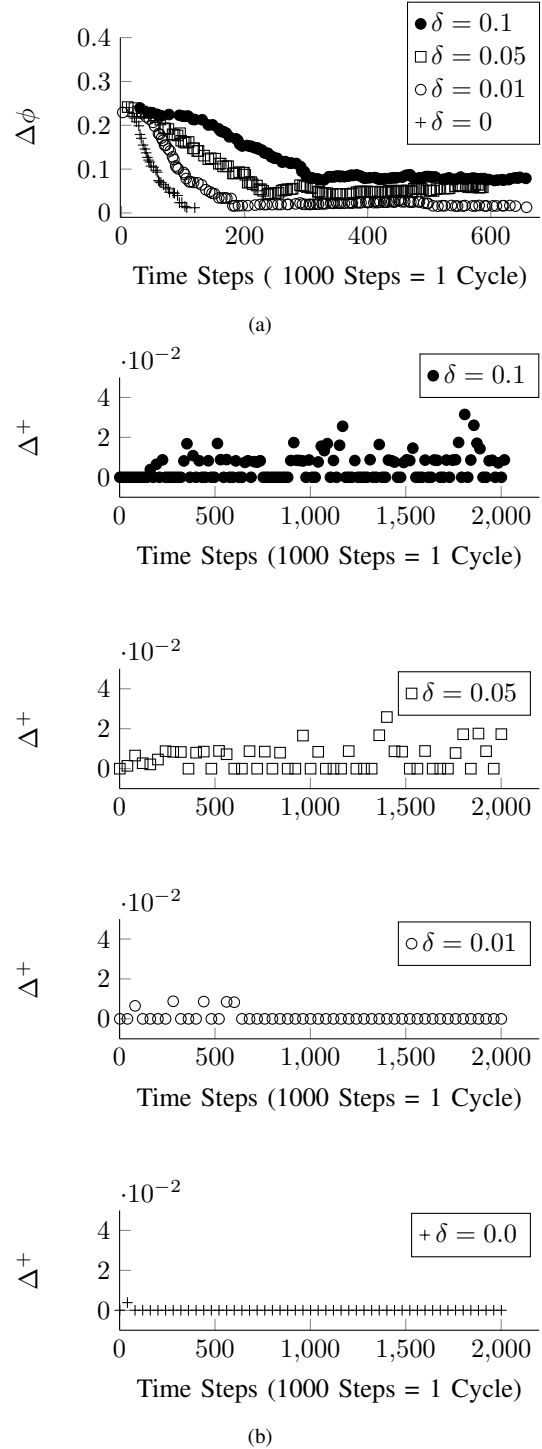}
   }
      \subfigure[]{
     \setlength\figureheight{0.2\textwidth}
   \setlength\figurewidth{0.7\textwidth}
%
\begin{tikzpicture}

\begin{axis}[%
width=\figurewidth,
height=\figureheight,
scale only axis,
xmin=0,
xmax=2200,
ymin=-0.01,
ymax=0.05,
ylabel={$\Delta^{+}$},
xlabel = {Time Steps (1000 Steps = 1 Cycle)},
name=plot1,
axis x line*=bottom,
axis y line*=left,
legend style={draw=black,fill=white,legend cell align=left,at={(1,1.2)}}
]
\addplot[only marks,mark=*,mark options={},color=black] plot table[row sep=crcr,]{%
1	0\\
17	0\\
33	0\\
49	0\\
65	0\\
81	0\\
97	0\\
113	0\\
129	0\\
145	0\\
161	0.0038807\\
177	0\\
193	0.0065041\\
209	0\\
225	0.008730413\\
241	0\\
257	0\\
273	0\\
289	0\\
305	0\\
321	0\\
337	0.00819512\\
353	0.0168091585\\
369	0\\
385	0.010829707\\
401	0\\
417	0.00826609409\\
433	0\\
449	0\\
465	0.008273627\\
481	0\\
497	0\\
513	0.01692750508\\
529	0.00882846161\\
545	0\\
561	0.00885534605\\
577	0\\
593	0\\
609	0.0082157575796\\
625	0\\
641	0\\
657	0.00745427528\\
673	0.00826623665\\
689	0\\
705	0.0078379204562\\
721	0.0075467320631\\
737	0.0082808256266\\
753	0\\
769	0\\
785	0\\
801	0\\
817	0\\
833	0\\
849	0\\
865	0\\
881	0\\
897	0.00838890825627\\
913	0.0173752232481\\
929	0.00849171469952\\
945	0.00840730012609\\
961	0.00818000211362\\
977	0\\
993	0.0086507745191\\
1009	0\\
1025	0\\
1041	0.00767227357344\\
1057	0.0155614265014\\
1073	0.0133408030237\\
1089	0.0168562227306\\
1105	0\\
1121	0.00841787261124\\
1137	0\\
1153	0.0160281979131\\
1169	0.0255425925724\\
1185	0\\
1201	0\\
1217	0\\
1233	0.00838079407252\\
1249	0\\
1265	0\\
1281	0\\
1297	0\\
1313	0.00880927872239\\
1329	0\\
1345	0\\
1361	0.01642682601\\
1377	0\\
1393	0\\
1409	0.00861858201453\\
1425	0.00831089124783\\
1441	0.00762100377533\\
1457	0\\
1473	0\\
1489	0.0073952452879\\
1505	0.00874360664288\\
1521	0.00840031601073\\
1537	0.0145906283887\\
1553	0\\
1569	0\\
1585	0\\
1601	0.00859647343082\\
1617	0\\
1633	0\\
1649	0.00867775720421\\
1665	0.00847320627993\\
1681	0\\
1697	0\\
1713	0.0086132710186\\
1729	0\\
1745	0.00858772886874\\
1761	0.00881628843753\\
1777	0.0174345307018\\
1793	0\\
1809	0.0314716400129\\
1825	0\\
1841	0.00863825742355\\
1857	0.0260482711627\\
1873	0.0170552367395\\
1889	0.0143252593505\\
1905	0\\
1921	0\\
1937	0.00826774748919\\
1953	0.00854832415019\\
1969	0\\
1985	0.00720927093074\\
2001	0\\
2017	0.00870576465431\\
};
\addlegendentry{{$\delta = 0.1$}};

\end{axis}

\begin{axis}[%
width=\figurewidth,
height=\figureheight,
yshift=-0.8cm,
scale only axis,
xmin=0,
xmax=2200,
ymin=-0.01,
ymax=0.05,
ylabel={$\Delta^{+}$},
name=plot2,
at=(plot1.below south west),
anchor=above north west,
axis x line*=bottom,
axis y line*=left,
legend style={draw=black,fill=white,legend cell align=left}
]
\addplot[only marks,mark=square,mark options={},color=black] plot table[row sep=crcr,]{%
1	0\\
41	0.0014156\\
81	0.0065735\\
121	0.0027323\\
161	0.0021482\\
201	0.0045935\\
241	0.00877097924\\
281	0.008428661\\
321	0.00834328241\\
361	0\\
401	0.008010287\\
441	0.00855535823\\
481	0\\
521	0.0088897877243\\
561	0.00722580704\\
601	0\\
641	0\\
681	0.0087743580704\\
721	0\\
761	0.008551244087\\
801	0\\
841	0.00801462415699\\
881	0\\
921	0\\
961	0.0165714041597\\
1001	0\\
1041	0.00845278212441\\
1081	0\\
1121	0\\
1161	0\\
1201	0.00880584534432\\
1241	0\\
1281	0\\
1321	0\\
1361	0.0167447276036\\
1401	0.0258533403408\\
1441	0.00870849891131\\
1481	0.00853257882474\\
1521	0\\
1561	0\\
1601	0.00885327476077\\
1641	0\\
1681	0\\
1721	0\\
1761	0.00785009819679\\
1801	0.0172380908261\\
1841	0\\
1881	0.0175987614508\\
1921	0.00876565356153\\
1961	0\\
2001	0.0173046977388\\
};
\addlegendentry{{$\delta = 0.05$}};

\end{axis}
\begin{axis}[%
width=\figurewidth,
height=\figureheight,
yshift=-0.8cm,
scale only axis,
xmin=0,
xmax=2200,
ymin=-0.01,
ymax=0.05,
ylabel={$\Delta^{+}$},
xlabel = {Time Steps (1000 Steps = 1 Cycle)},
name=plot3,
at=(plot2.below south west),
anchor=above north west,
axis x line*=bottom,
axis y line*=left,
legend style={draw=black,fill=white,legend cell align=left}
]
\addplot[only marks,mark=o,mark options={},color=black] plot table[row sep=crcr,]{%
1	0\\
41	0\\
81	0.0065239\\
121	0\\
161	0\\
201	0\\
241	0\\
281	0.00879584309\\
321	0\\
361	0\\
401	0\\
441	0.008654084\\
481	0\\
521	0\\
561	0.00851001128\\
601	0.00835063514\\
641	0\\
681	0\\
721	0\\
761	0\\
801	0\\
841	0\\
881	0\\
921	0\\
961	0\\
1001	0\\
1041	0\\
1081	0\\
1121	0\\
1161	0\\
1201	0\\
1241	0\\
1281	0\\
1321	0\\
1361	0\\
1401	0\\
1441	0\\
1481	0\\
1521	0\\
1561	0\\
1601	0\\
1641	0\\
1681	0\\
1721	0\\
1761	0\\
1801	0\\
1841	0\\
1881	0\\
1921	0\\
1961	0\\
2001	0\\
};
\addlegendentry{{$\delta = 0.01$}};

\end{axis}

\begin{axis}[%
width=\figurewidth,
height=\figureheight,
yshift=-0.8cm,
scale only axis,
xmin=0,
xmax=2200,
ymin=-0.01,
ymax=0.05,
ylabel={$\Delta^{+}$},
xlabel = {Time Steps (1000 Steps = 1 Cycle)},
name=plot4,
at=(plot3.below south west),
anchor=above north west,
axis x line*=bottom,
axis y line*=left,
legend style={draw=black,fill=white,legend cell align=left}
]
\addplot[only marks,mark=+,mark options={},color=black] plot table[row sep=crcr,]{%
1	0\\
41	0.0038213\\
81	0\\
121	0\\
161	0\\
201	0\\
241	0\\
281	0\\
321	0\\
361	0\\
401	0\\
441	0\\
481	0\\
521	0\\
561	0\\
601	0\\
641	0\\
681	0\\
721	0\\
761	0\\
801	0\\
841	0\\
881	0\\
921	0\\
961	0\\
1001	0\\
1041	0\\
1081	0\\
1121	0\\
1161	0\\
1201	0\\
1241	0\\
1281	0\\
1321	0\\
1361	0\\
1401	0\\
1441	0\\
1481	0\\
1521	0\\
1561	0\\
1601	0\\
1641	0\\
1681	0\\
1721	0\\
1761	0\\
1801	0\\
1841	0\\
1881	0\\
1921	0\\
1961	0\\
2001	0\\
};
\addlegendentry{{$\delta = 0.0$}};

\end{axis}

\end{tikzpicture}%
   }
  
  \caption{ Convergence behavior of EBS with different values of $\delta$ when  $\sigma = 0.01$ when $\varepsilon = 0.1$.  Fig (a) shows that value of $\delta$ increases, the average phase difference $\Delta \phi$ also increases and meets the boundary condition shown in Equation~(\ref{Eq-delay}).  Fig (b) shows that average phase advancement converges to zero when $\delta = 0.0$ and $0.01$ (represents EBS convergence) and stays to a fluctuating value when  $\delta = 0.1$ and $0.05$, hence represents EBS non-convergence.}
\label{fig:figure2}
\end{minipage}
\end{figure}

When all the delays are deterministic and satisfy the above Equation~(\ref{Eq-delay}), then the system converges within the given  precision bounds as shown in Figure~\ref{fig:figure2}. 
In Figure~\ref{fig:figure2}~(a) we present the convergence behavior of EBS for various values of $\delta$ when $\varepsilon = 0.1$ and $\sigma = 0.01$.  We observe that, as the value of $\delta$ increases, the average phase difference $\Delta \phi$ also increases and meets the boundary condition shown in Equation~(\ref{Eq-delay}). The Figures~\ref{fig:figure2}~(b) shows the  value of average phase advancement that converges to zero when $\delta = 0.0$ and $0.01$, which represents EBS convergence whereas represents EBS non-convergence behavior when the average phase advancement value keeps fluctuating between $0$ and $0.04$ at  $\delta = 0.1$ and $0.05$.

Our numerical experiments also suggest that $\varepsilon > 2\delta$ is required to achieve stability. Moreover, if the stability criterion is satisfied, higher values of $\sigma$ leads an improvement of the precision of  synchronization. Hence, we study and derive the relationship between $\sigma$, $\delta$ and  $\varepsilon$ in the next Section~\ref{EBSpara}.

\section{EBS Configuration Parameters}\label{EBSpara}
The performance of the EBS scheme depends on the values of its three configuration parameters: $\varepsilon$, $\sigma$ and $S_{Th}$. 
These parameter values reflect the network density and the uniformity of the nodes in terms of degree of connectivity in the network. For a given average degree connectivity per node in 
the network, we analyze  $\varepsilon$, $\sigma$, and $S_{Th}$, respectively.  \\

\noindent
{\bf Effects of $\mathbf{\varepsilon}$ -- }  The $\varepsilon$ parameter plays an important role in the EBS scheme because it is directly proportional to the duty-cycle, that is, the awake period of a node corresponding to its SETW.  In a DCM network, we wish to keep energy consumption of a node to minimum by maintaining a small value of  $\varepsilon$ and SETW. 

\begin{figure}[ht!] 
   \centering
   \includegraphics[width=2.5in]{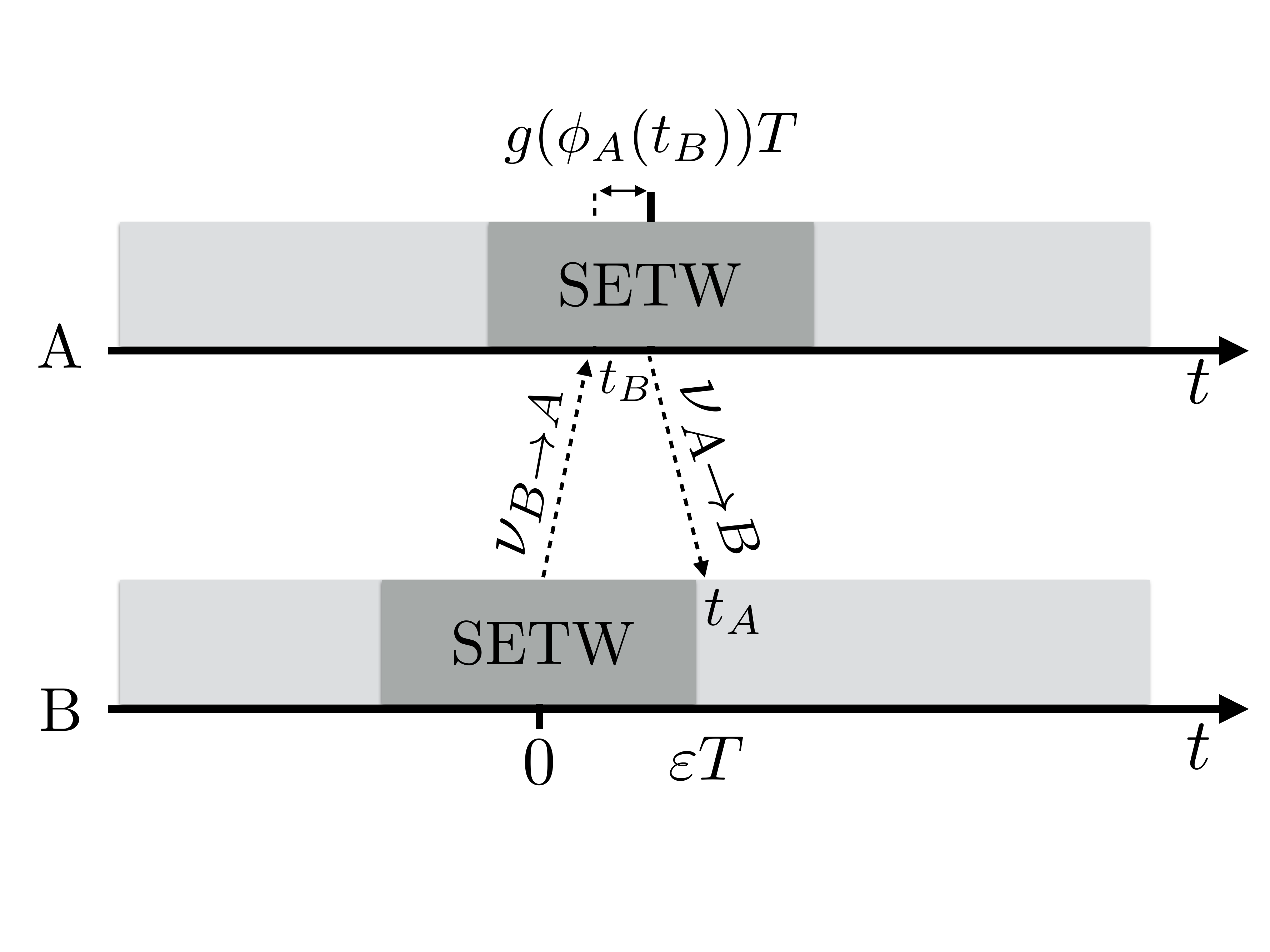} 
\caption{Overshooting condition occurs when node B receives broadcast message from node A at its time $t_A$, which is out of its SETW.}
\label{fig:EBSsvg4}
\end{figure}

However, regarding the convergence of the EBS scheme, the size of the SETW plays an inverse role, that is, the larger the $\varepsilon$  and  SETW,  the quicker the EBS scheme converges. Therefore, a trade-off is required to find the minimum size of SETW that can guarantee fast convergence without the problem of overshooting as shown in Figure~\ref{fig:EBSsvg4}.

The overshooting situation occurs when node A receives a broadcast message from node B after a time-delay $(\nu_{B\rightarrow A})$, see Figure~\ref{fig:EBSsvg4}. 
This overshooting situation leads to a chasing condition, both nodes will keep on updating their phases. 
To avoid this overshooting condition, we can limit the lower bound of the $\varepsilon T$ to:
\begin{equation}\label{eqSETW}
\varepsilon  T  \geq (\nu_{A\rightarrow B}) + g(\phi(t_B))T +  (\nu_{B\rightarrow A}).
\end{equation}
If $\nu_{A\rightarrow B} \approx \nu_{B\rightarrow A} = \nu$,  Equation~(\ref{eqSETW}) can be rewritten as:\\
$ \varepsilon \geq  g(\phi(t_B))T +  2\nu)/T$. Notice that in this construction $\varepsilon \leq 0.5$. 
Therefore, we obtain:
\begin{equation}\label{eqSEToo}
 0.5 \geq \varepsilon \geq g(\phi(t_B)) +  \frac{2\nu}{T}.
\end{equation}
So, even if the nodes are synchronized $g(\phi(t_B)) = 0$ (when we consider the immediate transmission of broadcast messages), 
the lower value of $\varepsilon$ is bounded by $\frac{2\nu}{T} \approx 2\delta$. The optimal value of $\varepsilon$ is chosen by considering the duty-cycle requirements and density of the network. 

For example, we let $\beta$ be the maximum time a node needs to receive and process a broadcast message. Assuming that each neighbourhood node sends a single message in $T$. So, the maximum 
time a node is required to be awake to receive and process messages from its neighbourhood of size $|N|$ is $\beta |N|$; thus the size of SETW should be $|SETW| = \beta |N|  + 4\nu$, which yields $2 \varepsilon  T \geq \beta |N| + 4\nu$. So, the optimal value of $\varepsilon$ reads as:

\begin{equation}\label{eqSETss}
\varepsilon_{opt} = \frac{\beta |N| + 4\nu}{2T}.
\end{equation} \\

\noindent
{\bf Effects of  phase advancement function -- } Our coupling function is linear and $\sigma$ defines the rate of convergence in the \textit{initialization phase}.  Recall that a large value of $\sigma$ slows down the convergence of the EBS scheme. For faster synchronization, we must choose smaller values of $\sigma$. Notice that we do not use $\sigma$ = 0, because this condition leads to message collisions in the network\footnote{A node immediately transmits its broadcast message, and if all the nodes in the local neighbourhood  do the same, it leads to message collisions.}. The maximum value of the coupling $\sigma$ can be obtained by performing a similar analysis as in Equation~(\ref{StabC}). In the presence of transmission delays we obtain: 
\begin{equation}\label{eqSETqq}
\sigma_{max}  \leq \frac{\varepsilon - \frac{2\nu}{T}}{ (1-\varepsilon)}. 
\end{equation}\\

\section{EBS Test-bed Implementation Details}\label{EBS-Imp}
To observe real-world behaviors, we implemented the EBS scheme on a test-bed. We used the UPMA~\cite{MLA2007} framework in TinyOS 2.x for the CC2420 radio, compliant to \textit{IEEE 802.15.4} standards, with a data rate of $250$ kbps. The UPMA framework is built on the  CSMA MAC, and we chose to couple the EBS components to the MAC layer instead of the application layer to avoid buffer delays caused by the intermediate stack queues and buffers. We implemented two versions of the EBS scheme to understand its behavior. The two implementations are:  (a) EBS without Reach-back\footnote{Recall reach-back allows a node to keep a record of the phase advancement of the current period and advances its phase immediately at the start of the next time period instead of firing immediately at current period.} to enable stand-alone synchronization support in the absence of upper-layer broadcast messages, and (b) EBS with partial Reach-back in which the broadcast messages are those generated from an application or routing layer, and are piggybacked with a sync-byte representing the EBS broadcast messages.  

\subsection{EBS without Reach-back}\label{EBS-RB}

In the latter implementation of the EBS scheme, when a node receives the broadcast-message from other nodes,  it advances its phase according to the EBS phase advancement function Equation~(\ref{eq2}). 
The node  broadcast the messages each time when $\phi(t) = 1$.  This means, if there is a message pending to be transmitted at $\phi(t) = 1$, EBS  piggybacks this message with its sync-byte and transmits it as the EBS broadcast-message. If there are no pending messages at the time, then the EBS layer creates a new  synchronization message and broadcast it.

The advantage of this scheme is that nodes converge quickly in the \textit{synchronization state} by exchanging messages without the need for \textit{reach-back}. However, this scheme has a high cost when overshooting occurs in \textit{synchronization state}. The overshooting occurs when $\varepsilon < 2\delta$  or $\varepsilon < (\nu_{A\rightarrow B} + \nu_{B\rightarrow A} )$, as previously derived  in Section~\ref{DeterDelay}.
Here, the nodes chase each other's broadcast messages without achieving stable synchronization. As a result, the whole network becomes congested with EBS broadcast messages, the network may never go to \textit{synchronization state} irrespective of $S_{Th}$. If overshooting occurs temporarily in \textit{steady duty-cycled state},  for lower $S_{Th}$, the network  flips back to the \textit{synchronization state}.

\subsection{EBS with Partial Reach-back}\label{EBS-PRB}

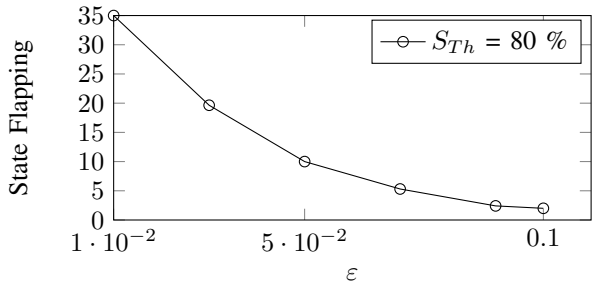
\begin{figure}[ht!]
\centering
  \setlength\figureheight{0.15\textwidth}
   \setlength\figurewidth{0.35\textwidth}
%
\begin{tikzpicture}

\begin{axis}[%
width=\figurewidth,
height=\figureheight,
scale only axis,
xmin=0.01,
xmax=0.11,
xtick={0.01,  0.05,  0.1},
xlabel={$\varepsilon$},
ymin=0,
ymax=35,
ytick={ 0,  5, 10, 15, 20, 25, 30, 35, 40, 45, 50},
ylabel={State Flapping},
legend style={draw=black,fill=white,legend cell align=left}
]
\addplot [color=black,solid,mark=o,mark options={solid}]
  table[row sep=crcr]{%
0.01	35\\
0.03	19.6441268692716\\
0.05	10\\
0.07	5.30845730824891\\
0.09	2.41521273516642\\
0.1	2\\
};
\addlegendentry{{$S_{Th}$ = 80 \%}};

\end{axis}
\end{tikzpicture}%
    \caption{Individual behavior of a node, which has average connectivity 20 in Motelab Test-bed, where all nodes running EBS with reach-back.}
  \label{fig:EBSflip1}
\end{figure}

 The EBS scheme is implemented at the MAC layer; the EBS layer receives a broadcast message from the application or routing layer and transmits it after piggy-backing as EBS broadcast message.
To establish coordinated SETWs during the \textit{synchronization state}, a node advances its phase but transmits broadcast messages only when it has a new schedule broadcast message. The advantage of this scheme is that when overshooting occurs this scheme avoids the network flooding with broadcast messages, unlike the above implementation. However, this leads to nodes state flapping between
\textit{synchronization} and the \textit{duty-cycled states}, shown experimentally in Figure~\ref{fig:EBSflip1}. 
 In Figure~\ref{fig:EBSflip1}, $T =10$~s and we fixed $\varepsilon = 0.01$ that is smaller than the  $\varepsilon_{opt} \approx 0.04 $ derived from Equation~(\ref{eqSETss}), and a node flaps between synchronized and non-synchronized modes. That is, when a node is in its \textit{synchronization state}, it switches to a \textit{duty-cycle state}, however, in the non-synchronized mode,  it returns to the $100\%$ duty-cycled mode again. Further, when $\varepsilon = 0.05$, then the flapping reduces significantly as predicted by our theory.

 \subsection{Neighbourhood Size}\label{EBS-FRB}
 For each node to understand its degree of synchronicity, the EBS implementation requires information regarding the neighbourhood size to calculate a node's $S_{Th}$. One approach is to get the neighbourhood size at the start by keeping  all nodes awake for the initial few periods of $T$ and identifying its neighbours via overhearing. To make the EBS scheme more agile, it is better to update neighbourhood size periodically during the  EBS \textit{duty-cycled state}. Nevertheless, in the implementation presented in this paper, EBS does not maintain its own neighbourhood table but uses the neighbourhood size, calculated from the number of broadcast messages overheard in a single broadcast message interval $T$ while it was in the \textit{initialization state}. Alternatively, link estimation mechanisms found in many routing protocols require neighbourhood information so this  can be freely used to make a precise neighbourhood size estimation.

\section{Experimental Setup and Results}\label{EBS-setup}\label{EBS-Results}
We perform our experiments on two different size test-beds: Motelab ($87$ Telosb nodes)~\cite{Motelab2010} and a smaller $10$ nodes MicaZ testbed~\cite{MicaZ}.  We begin our experiments by fixing the value of the broadcast time interval $T$ and varying the synchronization threshold $S_{Th}$, then we measure the duty-cycle of the network after every second interval.  We present the results of the experiments executed on Motelab, with a broadcast interval time $T$ set at $10$, $20$, and $30$ seconds (representing typical routing control message intervals).  Recall the SETW width varies according to the Equation~(\ref{eqSETss}).
 
 For each run (either varying $T$ or $S_{Th}$), we measured the duty-cycle of each node and their wake-up times, as a percentage. In the resulting graphs, we present the average duty-cycle and average throughput 
of the network by varying the $S_{Th}$ from $95$ to $20$. The average duty cycle is calculated by keeping a internal timer for each node, which keeps record of  wake-up slots during the run time of the experiment.  
We then use average duty cycle of each node to calculate the network-wide average duty cycle.  For broadcast communication, we use the following Equation to calculate the throughput percentage in one time period:\\
$$
Throughput = \frac{\mbox{Sum of all broadcast messages received}}{\mbox{network average degree} \times n } \times 100 \\
$$
The average network-wise degree of connectivity is calculated by sum of all the messages received divided by total number of nodes in the $100\%$ duty-cycled mode.  The EBS scheme exchanges broadcast messages in SETW, for unicast messages, it can easily be integrated with unicast message handling MAC layer~\cite{pyDcoss2011}. \\
In the next Section, we present EBS performance results showing both convergence parameter effects and performance and compare with the nearest state-of-the-art slot management protocol. 

\subsection{Calculation of Deterministic Propagation Delay}\label{RBS-calculation}
From Equation~(\ref{eqSETss}), it is clear that the value of $\varepsilon_{opt}$ is proportional to propagation and processing delays $\beta$. The value of $\beta$ is non-deterministic in real environments, however the minimum value ($C_{0}\leq \beta$)  as a constant parameter that is required in EBS scheme  to  calculate size of SETW to ensure convergence in the \textit{synchronization  state}. To get approximate value of $C_{0}$, we first performed our initial experiments to observe the behavior of the EBS with a prefixed 
$\varepsilon$ varying from $0.01$ to $0.1$. 
We found that nodes flapping with a low value of $\varepsilon$ are more prominent compared to large values of $\varepsilon$, shown in Figure~\ref{fig:EBSflip1}.  

We ran initial experiments to derive the value of $C_{0}$, we  found it to be $C_{0}\geq 50$~ms for the network to converge when the average connectivity of this network was $2-3$ nodes. However, in a non-uniform network, where some nodes have high degree of connectivity, the EBS calculates  the adaptive value $C^{i}$ of node $i$ that is necessary for  achieving synchronization using constant value $C_{0}$  and the node's neighbourhood density and other factors that can be further explored. However, for the experiments in this paper, we choose to implement EBS~(without \textit{reach-back} implementation) 
with adaptive $C^{i}$ given as follows:
\begin{equation}\label{eqSETvv}
\begin{split}
C^{i}&=(C_{0}\times |N_i| \times S_{Th}/100)~ms \\
&=(50\times |N_i| \times S_{Th}/100)~ms
\end{split}
\end{equation}
Note that the initial experiments shown in Figure~\ref{fig:EBSflip1}, are performed taking into account $\sigma_{max}  \leq \frac{\varepsilon}{(1-\varepsilon)}$ and by assuming $2\nu/T = 0$ in Equation~(\ref{eqSETqq}).
\begin{figure}[ht!]
   \centering
    \setlength\figureheight{0.20\textwidth}
     \setlength\figurewidth{0.35\textwidth}
     \mbox{
       \subfigure[]{\scalebox{1.0}{\includegraphics[width=8cm, height = 6cm]{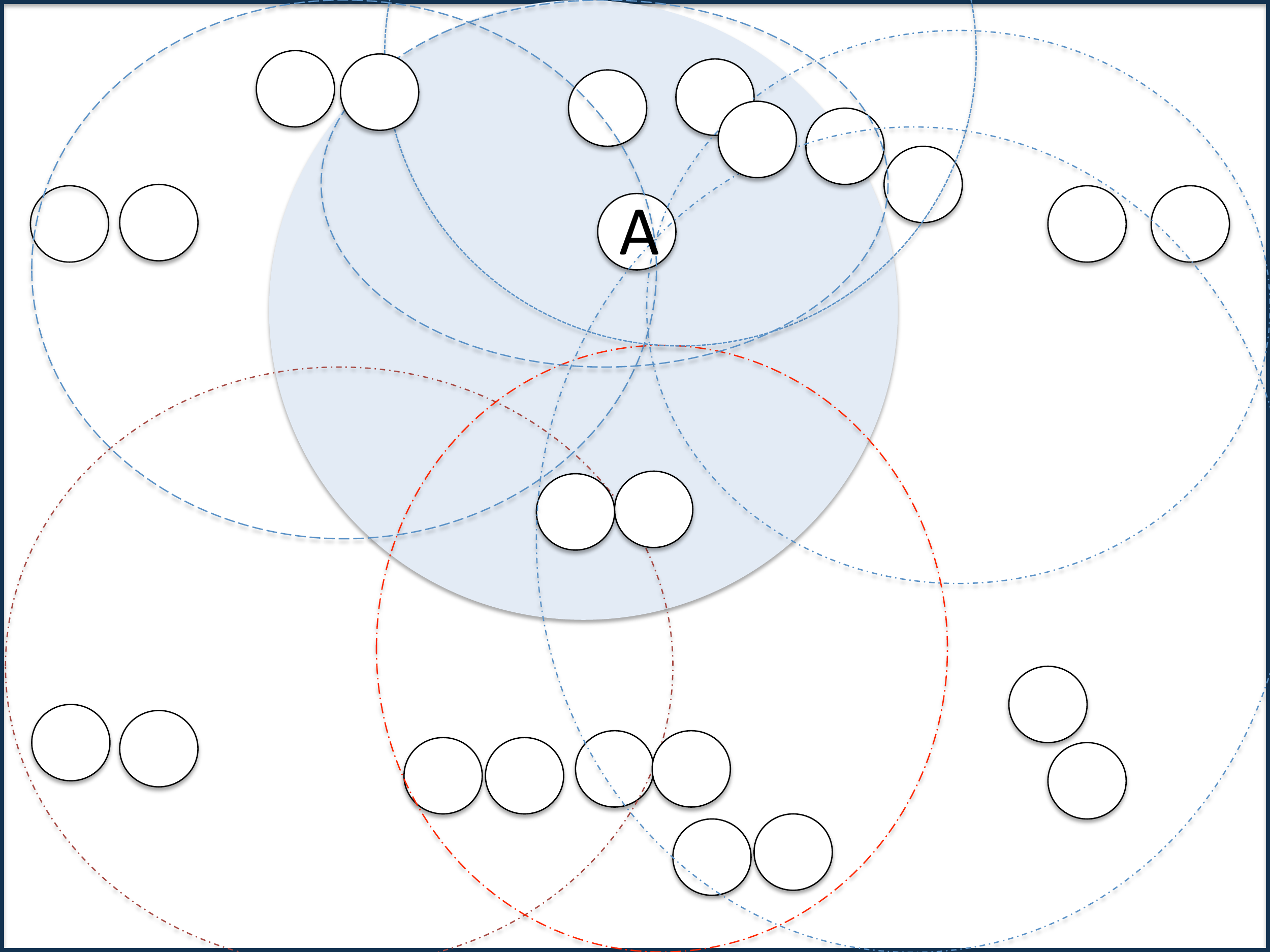}}} \quad               
         } 
    \mbox{      
     \subfigure[]{\scalebox{1.0}{ 
%
\begin{tikzpicture}

\begin{axis}[%
width=\figurewidth,
height=\figureheight,
scale only axis,
xmin=20,
xmax=95,
xtick={20, 30, 40, 50, 60, 70, 80, 90},
ymin=0,
ymax=10,
ytick={ 0,  2,  4,  6,  8, 10},
ylabel={Duty-Cycle (\%)},
name=plot1,
legend style={draw=black,fill=white,legend cell align=left}
]
\addplot [color=black,solid,mark=square,mark options={solid}]
  table[row sep=crcr]{%
20	0.5\\
25	0.5\\
30	0.5\\
35	0.5\\
40	0.5\\
45	0.515625\\
50	0.55\\
55	0.584375\\
60	0.6\\
65	0.6\\
70	0.6\\
75	0.6\\
80	0.6\\
85	0.738095238095238\\
90	1.07619047619048\\
95	1.5\\
};
\addlegendentry{Mean, C = $C_{0}$ = 50 ms};

\addplot [color=black,dashed]
  table[row sep=crcr]{%
20	1\\
25	0.809375\\
30	0.65\\
35	0.540625\\
40	0.5\\
45	0.515625\\
50	0.55\\
55	0.584375\\
60	0.6\\
65	0.6\\
70	0.6\\
75	0.6\\
80	0.6\\
85	1.35185185185185\\
90	3.19259259259259\\
95	5.5\\
};
\addlegendentry{Node A, C = $C_{0}$ = 50 ms};

\addplot [color=black,solid,mark=o,mark options={solid}]
  table[row sep=crcr]{%
20	1.7\\
25	1.846875\\
30	1.975\\
35	2.065625\\
40	2.1\\
45	2.1\\
50	2.1\\
55	2.1\\
60	2.1\\
65	2.13289473684211\\
70	2.22105263157895\\
75	2.34868421052632\\
80	2.5\\
85	2.73313840155945\\
90	3.08323586744639\\
95	3.5\\
};
\addlegendentry{Mean,  C $>$ 50 ms};

\end{axis}

\begin{axis}[%
width=\figurewidth,
height=\figureheight,
scale only axis,
xmin=20,
xmax=95,
xtick={20, 30, 40, 50, 60, 70, 80, 90},
xlabel={$S_{Th}$},
ymin=40,
ymax=100,
ytick={ 40,  50,  60,  70,  80,  90, 100},
ylabel={Throughput (\%)},
at=(plot1.below south west),
anchor=above north west
]
\addplot [color=black,solid,mark=square,mark options={solid},forget plot]
  table[row sep=crcr]{%
20	60\\
25	51.25\\
30	45\\
35	41.25\\
40	40\\
45	43.125\\
50	50\\
55	56.875\\
60	60\\
65	60\\
70	60\\
75	60\\
80	60\\
85	60\\
90	60\\
95	60\\
};
\addplot [color=black,dashed,forget plot]
  table[row sep=crcr]{%
20	61\\
25	49.421875\\
30	43.125\\
35	40.515625\\
40	40\\
45	48.2441737288136\\
50	66.5677966101695\\
55	85.3575211864407\\
60	95\\
65	96.6737288135593\\
70	97.9322033898305\\
75	98.7245762711864\\
80	99\\
85	90.8095238095238\\
90	70.6190476190476\\
95	45\\
};
\addplot [color=black,solid,mark=o,mark options={solid},forget plot]
  table[row sep=crcr]{%
20	65\\
25	69.4561820652174\\
30	73.508152173913\\
35	77.0560461956522\\
40	80\\
45	82.4673913043478\\
50	84.6376811594203\\
55	86.4891304347826\\
60	88\\
65	89.1209677419355\\
70	89.989247311828\\
75	90.8629032258064\\
80	92\\
85	93.6180235535074\\
90	95.6661546338966\\
95	98\\
};
\end{axis}
\end{tikzpicture}%
}}\quad         
          }
\caption{ (a) The Motelab test-bed topology where nodes are shown as small circles with radio coverage shown by large circles;  (b) we study the behavior of individual node A (radio coverage of node A shown  by the shaded circle) in the given topology, where all nodes run EBS with 
different synchronization thresholds~($S_{Th}$);  we also analyze the average duty-cycle and average throughput percentages of all nodes when $C = C_{0} = 50$~ms and when $C$ is adaptive according to Equation~(\ref{eqSETvv}).}
\label{Resultssth}
 \end{figure}
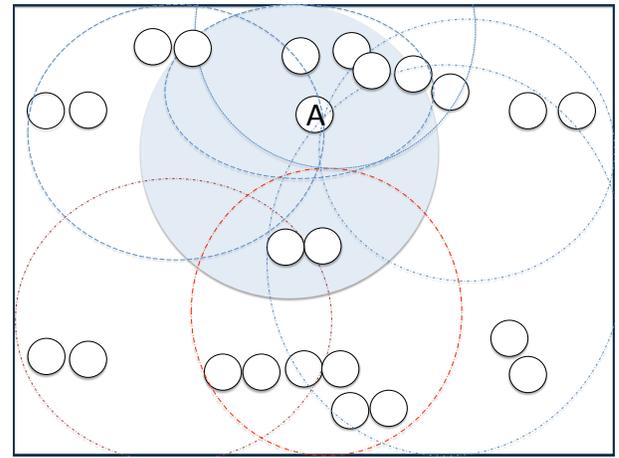
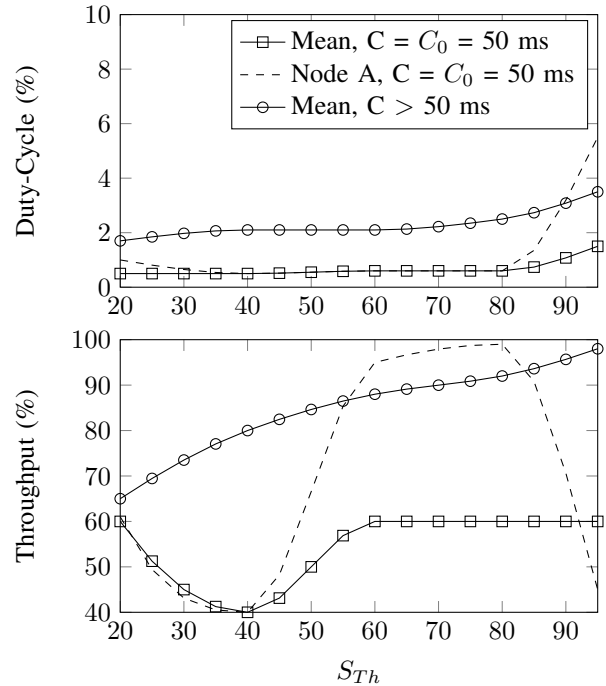 
\subsection{Effects of Synchronization Threshold $S_{Th}$}\label{EBS-STH}
We study the effect of the Synchronization Threshold on the nodes' convergence in the \textit{synchronization state}. 
For these experiments, we set $T = 30$~s and vary the synchronization threshold $S_{Th}$ from $20$ to $95$, the results  are shown in Figure~\ref{Resultssth}.

Figure~\ref{Resultssth} (a) shows the Motelab~\cite{Motelab2010} topology on which we ran our experiments.  Figure~\ref{Resultssth}~(b) is composed of two figures, showing the comparative duty-cycle~($\%$) and throughputs~($\%$) achieved.

The figures present an average duty-cycle and throughput with  $C = C_{0} =  50$~ms for all nodes, node A's average duty cycle and throughput with $C = C_{0} =  50$~ms, and average duty-cycle  and throughput with adaptive $C$  which is calculated by each node using Equation~(\ref{eqSETvv}) of all nodes in the given topology. Due to high degree of connectivity, Node A goes out of synchronization for high value of $S_{Th} \geq 80$ when $C = C_{0} =  50$~ms, which results in decreased throughput  as shown in Figure ~\ref{Resultssth}~(b). To avoid these situations, adaptive $C$ is proposed. We found that adaptive $C$ follows a regular pattern regarding both duty-cycle and throughput results. Furthermore, adaptive $C$ improves the throughput significantly by $40\%$ at the small cost to the duty-cycle~($2-3\%$).  

We can also see that the duty-cycle of the node is nearly $100\%$ when $S_{Th} = 95$ as compared to other lower values of $S_{Th}$. This clearly shows that to achieve synchronicity with a high threshold the node must remain awake slightly longer as compared to the low threshold in the \textit{synchronization state}. In addition, note that, in this set of experiments, 
we perform neighbourhood size calculations only in the first $T$ period and we observe that due to lossy wireless links, this value might not be the actual neighbourhood size. For this reason, we selected the average number of neighbours by keeping nodes awake $100\%$ for the first 5$T$ in the following experiments.

\subsection{Effects of the Broadcast Message Interval, $T$}\label{EBS-T}
Figure~\ref{Resultsebstime} presents the effects of $T$ on the duty-cycle~($\%$) and throughput~($\%$). We use the three values of $T = 10$~s, $20$~s, $30$~s, and varied $S_{Th}$ from $20$ to $95$. We found that the duty-cycle increases nearly
 $0.1\%$ with every $10$ increase in $S_{Th}$ till $S_{Th} = 60$ for all values of $T$. However, for $S_{Th} > 60$, its incremental effect on the duty-cycle is  slightly higher ($1-3\%$) for low values of $T$, as compared 
to high values of $T$.
On the other hand, throughput is not much affected by $T$; higher values of $T$ perform better for low $S_{Th}$ and the  lower values of $T$ perform better for high $S_{Th}$. However, there is a gradual increase in throughput 
from $20-30\%$  to $80-90\%$ when $S_{Th}$ varies from $20$ to $95$, respectively.

\subsection{Comparative Performance}\label{EBS-Comparative}
We compare our result with \textit{Refractory Periods}~\cite{Degesys08} because it is the closest of the bio-inspired synchronization approaches to ours. To make it suitable for duty-cycling, we use the $T_{Ref} = T/2 $, as 
suggested in paper~\cite{Tyrrell10} for achieving quick loose synchronicity, we call this  the Modified Refractory Period~(MRF). 
 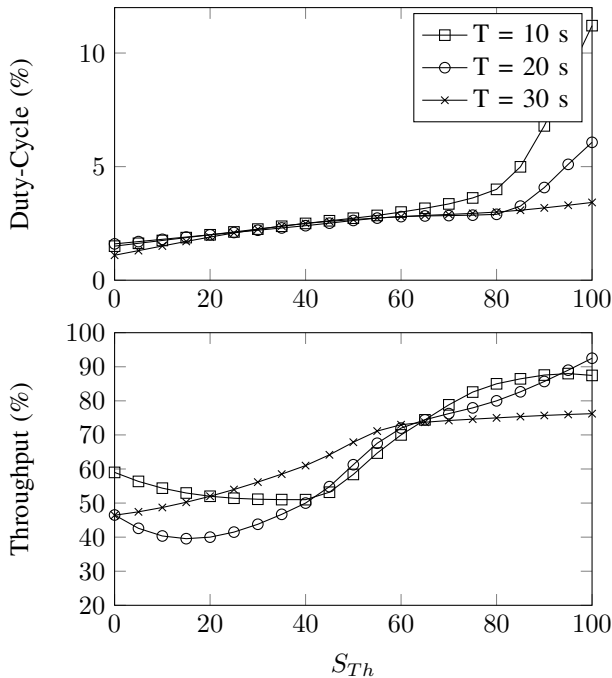
\begin{figure}[ht!]
\centering
  \setlength\figureheight{0.20\textwidth}
  \setlength\figurewidth{0.35\textwidth}
%
\begin{tikzpicture}

\begin{axis}[%
width=\figurewidth,
height=\figureheight,
scale only axis,
xmin=0,
xmax=100,
xtick={  0,  20,  40,  60,  80, 100},
ymin=0,
ymax=12,
ytick={ 0,  5, 10, 15},
ylabel={Duty-Cycle (\%)},
name=plot1,
legend style={draw=black,fill=white,legend cell align=left}
]
\addplot [color=black,solid,mark=square,mark options={solid}]
  table[row sep=crcr]{%
0	1.5\\
5	1.625\\
10	1.75\\
15	1.875\\
20	2\\
25	2.125\\
30	2.25\\
35	2.375\\
40	2.5\\
45	2.6171875\\
50	2.72916666666667\\
55	2.8515625\\
60	3\\
65	3.16550429184549\\
70	3.35801144492132\\
75	3.62151287553648\\
80	4\\
85	4.9912914140382\\
90	6.79326475463814\\
95	9\\
100	11.2055771283239\\
};
\addlegendentry{T = 10 s};

\addplot [color=black,solid,mark=o,mark options={solid}]
  table[row sep=crcr]{%
0	1.6\\
5	1.7\\
10	1.8\\
15	1.9\\
20	2\\
25	2.1\\
30	2.2\\
35	2.3\\
40	2.4\\
45	2.51125\\
50	2.63\\
55	2.73375\\
60	2.8\\
65	2.82863704819277\\
70	2.84469879518072\\
75	2.86341114457831\\
80	2.9\\
85	3.26243704978645\\
90	4.08002804870275\\
95	5.1\\
100	6.0695799069293\\
};
\addlegendentry{T = 20 s};

\addplot [color=black,solid,mark=x,mark options={solid}]
  table[row sep=crcr]{%
0	1.1\\
5	1.30234375\\
10	1.50625\\
15	1.70703125\\
20	1.9\\
25	2.08046875\\
30	2.24375\\
35	2.38515625\\
40	2.5\\
45	2.591875\\
50	2.67\\
55	2.738125\\
60	2.8\\
65	2.85229838709677\\
70	2.89612903225806\\
75	2.94189516129032\\
80	3\\
85	3.08090117767537\\
90	3.18330773169483\\
95	3.3\\
100	3.42375832053251\\
};
\addlegendentry{T = 30 s};

\end{axis}

\begin{axis}[%
width=\figurewidth,
height=\figureheight,
scale only axis,
xmin=0,
xmax=100,
xtick={  0,  20,  40,  60,  80, 100},
xlabel={$S_{Th}$},
ymin=20,
ymax=100,
ytick={  0,  10,  20,  30,  40,  50,  60,  70,  80,  90, 100},
ylabel={Throughput (\%)},
at=(plot1.below south west),
anchor=above north west,
legend style={draw=black,fill=white,legend cell align=left}
]
\addplot [color=black,solid,mark=square,mark options={solid}]
  table[row sep=crcr]{%
0	59\\
5	56.359375\\
10	54.375\\
15	52.953125\\
20	52\\
25	51.421875\\
30	51.125\\
35	51.015625\\
40	51\\
45	53.1829044117647\\
50	58.4044117647059\\
55	64.6737132352941\\
60	70\\
65	74.4131770086083\\
70	78.8272955523673\\
75	82.5777663199426\\
80	85\\
85	86.4607046070461\\
90	87.5636856368564\\
95	88\\
100	87.4607046070461\\
};

\addplot [color=black,solid,mark=o,mark options={solid}]
  table[row sep=crcr]{%
0	46.5\\
5	42.58984375\\
10	40.34375\\
15	39.55078125\\
20	40\\
25	41.48046875\\
30	43.78125\\
35	46.69140625\\
40	50\\
45	54.82109375\\
50	61.2520833333333\\
55	67.55703125\\
60	72\\
65	74.4456730769231\\
70	76.2551282051282\\
75	77.9370192307692\\
80	80\\
85	82.6483516483516\\
90	85.6813186813187\\
95	89\\
100	92.5054945054945\\
};

\addplot [color=black,solid,mark=x,mark options={solid}]
  table[row sep=crcr]{%
0	46.4285714285714\\
5	47.4296875\\
10	48.7053571428571\\
15	50.2354910714286\\
20	52\\
25	53.9787946428571\\
30	56.1517857142857\\
35	58.4988839285714\\
40	61\\
45	64.1607142857143\\
50	67.8571428571429\\
55	71.125\\
60	73\\
65	73.7203504043127\\
70	74.2304582210243\\
75	74.6253369272237\\
80	75\\
85	75.377158829989\\
90	75.712388938804\\
95	76\\
100	76.2343016871319\\
};

\end{axis}
\end{tikzpicture}%
   \caption{EBS duty-cycle and throughput with adaptive $C$.}
  \label{Resultsebstime} 
  \end{figure}
 
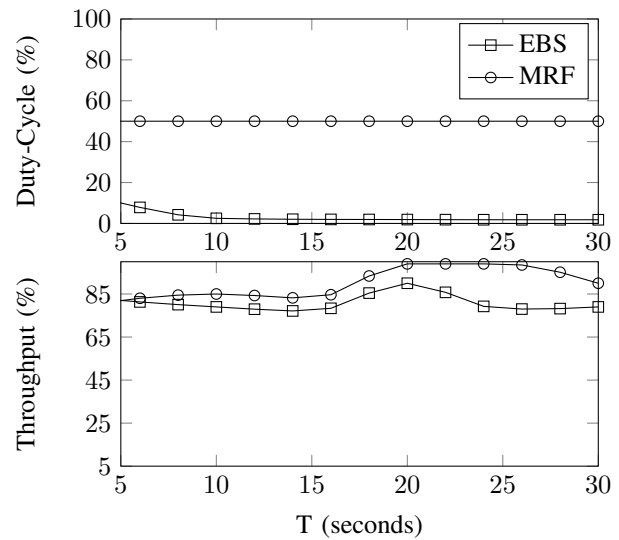
\begin{figure}[ht!]
\centering
  \setlength\figureheight{0.15\textwidth}
  \setlength\figurewidth{0.35\textwidth}
%
\begin{tikzpicture}

\begin{axis}[%
width=\figurewidth,
height=\figureheight,
scale only axis,
xmin=5,
xmax=30,
xtick={ 5, 10, 15, 20, 25, 30},
ymin=0,
ymax=100,
ytick={  0,  20,  40,  60,  80, 100},
ylabel={Duty-Cycle (\%)},
name=plot1,
legend style={draw=black,fill=white,legend cell align=left}
]
\addplot [color=black,solid,mark=square,mark options={solid}]
  table[row sep=crcr]{%
4	12.193\\
6	7.842\\
8	4.219\\
10	2.5\\
12	2.205\\
14	2.04333333333333\\
16	1.97146666666667\\
18	1.9336\\
20	1.9\\
22	1.8504\\
24	1.8072\\
26	1.8\\
28	1.8\\
30	1.8\\
};
\addlegendentry{EBS};

\addplot [color=black,solid,mark=o,mark options={solid}]
  table[row sep=crcr]{%
4	50\\
6	50\\
8	50\\
10	50\\
12	50\\
14	50\\
16	50\\
18	50\\
20	50\\
22	50\\
24	50\\
26	50\\
28	50\\
30	50\\
};
\addlegendentry{MRF};

\end{axis}

\begin{axis}[%
width=\figurewidth,
height=\figureheight,
scale only axis,
xmin=5,
xmax=30,
xtick={ 5, 10, 15, 20, 25, 30},
xlabel={T (seconds)},
ymin=5,
ymax=100,
ytick={ 5, 25, 45, 65, 85},
ylabel={Throughput (\%)},
at=(plot1.below south west),
anchor=above north west
]
\addplot [color=black,solid,mark=square,mark options={solid},forget plot]
  table[row sep=crcr]{%
4	82.7152\\
6	81.3168\\
8	80.0656\\
10	79\\
12	77.9504\\
14	77.1312\\
16	78.352\\
18	85.424\\
20	90\\
22	85.776\\
24	79.248\\
26	78.008\\
28	78.216\\
30	79\\
};
\addplot [color=black,solid,mark=o,mark options={solid},forget plot]
  table[row sep=crcr]{%
4	80.824\\
6	83.016\\
8	84.472\\
10	85\\
12	84.296\\
14	83.208\\
16	84.664\\
18	93.368\\
20	99\\
22	99\\
24	99\\
26	98.496\\
28	95.112\\
30	90\\
};
\end{axis}
\end{tikzpicture}%
      \caption{Comparative duty-cycle and throughput of EBS~(with adaptive $C$ and $S_{Th} = 80$) and MRF.}
 \label{Resultsebscomp} 
 \end{figure}

From Figure~\ref{Resultsebstime}, it is clear that for our current network 
topology, setting $S_{Th} = 80$  achieves more than $85\%$ throughput while keeping duty-cycle below $5\%$, which means saving nearly $80\%$ energy by avoiding idle listening.
We show that EBS achieves almost similar throughput, but with a large improvement in duty-cycling as shown in Figure~\ref{Resultsebscomp}.  To compare our results with MRF, 
we choose $S_{Th} = 80$.   

 In Figure~\ref{Resultsebscomp}, we use adaptive $C$ that is calculated by putting $C_{0}$ = 50~ms in Equation~(\ref{eqSETvv}), however, if we use value of  adaptive C calculated when  $C_{0}$ = 100~ms, we can achieve the same or higher throughput compared to MRF. Therefore, in this experiment we present the throughput achieved at the minimum duty-cycle for adaptive C (calculated using $C_{0}$ = 50~ms), which EBS can achieve while accommodating propagation and processing delays that are caused by the lower layers (e.g., the CSMA MAC layer). 
\section{Summary}\label{EBS-Conclusion}
In this paper, we presented EBS, a fully decentralized scheme for wireless embedded and IoT systems that ensures network-wide message broadcast in duty-cycled networks without requiring costly global synchronization. We have 
identified the challenges of providing such a mechanism, formalized the constraints therein and produced algorithms that are lightweight and robust to changing network topologies and failures.
EBS has been designed to provide support for broadcast messaging that uses local synchronization within a duty-cycled network. Further, EBS has been fully implemented on  a 
test-bed to show  the affects of convergence parameters and comparative performance with the nearest state-of-the-art protocol~(MRF). Here, we have showed that EBS achieves more than $80\%$ throughput while keeping the duty-cycle  down to less than $5\%$~(for broadcast message intervals, $T > 10$~s). Therefore, the results confirmed that EBS can match the throughput of comparative schemes that are continuously awake ($100\%$ duty-cycle) but with a considerable  reduction in the network-wide energy consumption.  
\section*{Acknowledgment}
This work was supported by UK-India Education and Research Initiative (UKIERI) award and United Kingdom NERC research grant (NE/I00694X/1). Tiago Periera was partially supported by FAPESP CeMEAI grant 2013/07375-0, Russian Science Foundation grant 14-41-00044 at the Lobayevsky University of Nizhny Novgorod and by the European Research Council - AdG grant number 339523 RGDD. 
\bibliographystyle{IEEEtran}
\bibliography{Yadav_July_2017.bbl}
\begin{IEEEbiography}[{\includegraphics[width=1in,height=1.25in,clip,keepaspectratio]{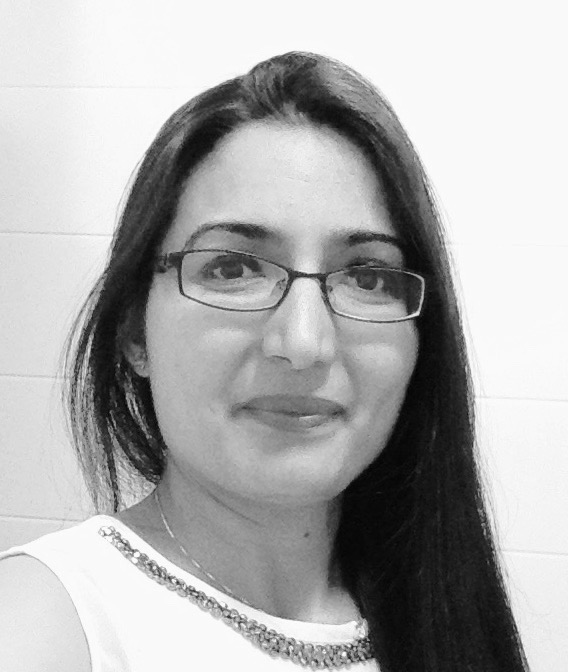}}]{Poonam Yadav} is a research and teaching associate at 
the Computer Laboratory, University of Cambridge.  She received the PhD degree in Computing from Imperial College London, in 2011 and M.Tech from IIIT, Allahabad, India, in 2007.  She is a recipient of UK-India Education and Research Initiative (UKIERI) PhD Award and has worked on various NERC, TSB, EU, EPSRC, IBM and Microsoft funded research projects and author of over 30 papers in Distributed Systems, Social Computing, Sensor Systems and IoT. She is currently the Chair of ACM-W UK professional Chapter and a member of ACM, IEEE and BCS.
\end{IEEEbiography}
\begin{IEEEbiography}[{\includegraphics[width=1in,height=1.25in,clip,keepaspectratio]{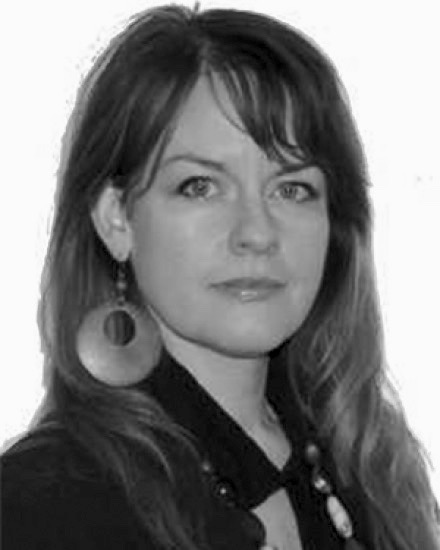}}]{Julie A. McCann}
is a professor in computer systems with Imperial College. Her research centers on highly decentralized and self-organizing scalable algorithms for spatial computing systems, e.g., wireless sensing networks. She leads both the Adaptive Embedded Systems Engineering Research Group and the Intel Collaborative Research Institute for Sustainable Cities and is currently working with NEC and others on substantive smart city projects. She has received significant funding from bodies such as the United Kingdom's EPSRC, TSB, and NERC as well as various international funds, and is an elected peer for the EPSRC. She has actively served on, and chaired, many conference committees and is currently associative editor of the ACM Transactions on Autonomous and Adaptive Systems. She is a fellow of the BCS.
\end{IEEEbiography}
\begin{IEEEbiography}[{\includegraphics[width=1in,height=1.25in,clip,keepaspectratio]{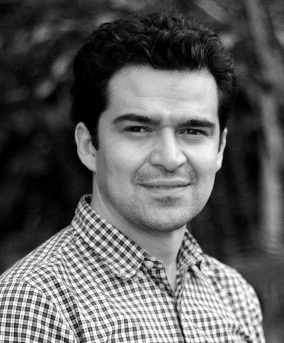}}] {Tiago  Pereira }
is a  professor of Mathematics at Institute of Mathematics and Computer Science, University of Sao Paulo, Brazil and author of over 40 papers on complex networks and dynamical systems. He is a principal investigator in the Center for Industrial Mathematics funded by FAPESP in partnership with leading companies. The main thrust of his research is the collective dynamics of complex systems concerning both theory and data-driven approaches. He obtained his PhD from Potsdam University in Germany, and held postdoctoral positions in Berlin, Sao Paulo and was a Leverhulme and Marie Curie Fellow at Imperial College London.
\end{IEEEbiography}
\enlargethispage{-2in}
\end{document}